\documentclass[a4paper,11pt]{article}
\pdfoutput=1 
\usepackage{jheppub}
\usepackage{algorithm}
\usepackage{algpseudocode}
\usepackage{amsmath}
\usepackage{xcolor}
\usepackage{epsfig}
\usepackage{graphicx}
\usepackage{appendix}
\usepackage{titlesec}
\usepackage{subfigure}

\def\A0#1{\Pi_{\rm #1}(0)}
\def\AP0#1{\Pi'_{\rm #1}(0)}
\def\be{\begin{equation}}
\def\ee{\end{equation}}
\def\bea{\begin{array}}
\def\eea{\end{array}}
\def\beqa{\begin{eqnarray}}
\def\eeqa{\end{eqnarray}}
\def\beqas{\begin{eqnarray*}}
\def\eeqas{\end{eqnarray*}}

\def\bp{\begin{picture}}
\def\ep{\end{picture}}
\def\bc{\begin{center}}
\def\ec{\end{center}}
\def\bfig{\begin{figure}}
\def\efig{\end{figure}}

\def\bit{\begin{itemize}}
\def\eit{\end{itemize}}
\def\nn{\nonumber}
\def\f{\frac}

\def\[{\left[}
\def\]{\right]}
\def\({\left(}
\def\){\right)}

\def\..{\left.}
\def\.{\right.}
\def\tl{\tilde}
\def\ra{\rightarrow}
\def\la{\leftarrow}

\def\tm{\times}

\def\la{\lambda}

\def\al{\alpha}

\def\ka{\kappa}

\def\ep{\epsilon}

\def\ga{\gamma}

\title{Positive definiteness constraints of effective scalar potential in Georgi-Machacek Model}
\author[\dagger~\S]{Xiaokang Du,}
\author[*]{Fei Wang$^1$\note{Corresponding author.}}
\affiliation[\dagger]{Institute of Physics, Henan Academy of Sciences, Zhengzhou 450046, P. R. China}
\affiliation[\S] {School of Physics, Henan Normal University, Xinxiang 453007, PR China}
\affiliation[*]{School of Physics, Zhengzhou University, Zhengzhou 450000, P. R. China}
\emailAdd{feiwang@zzu.edu.cn, xkdu@hnas.ac.cn}

\abstract{ The Georgi-Machacek (GM) Model extends the Higgs sector of the Standard Model by introducing additional triplets, preserving custodial symmetry at tree level and allowing large triplet vacuum expectation values (VEVs) of order $\cal{O}$(10) GeV. Theoretical constraints on the model's parameters include bounded-from-below~(BFB) conditions for the tree-level scalar potential. This study goes beyond the BFB constraints by examining the positive definiteness of the effective potential in the GM model to ensure the absence of deeper vacua in regions with large field values. Using a one-loop renormalization group-improved (RG-improved) effective potential and new criteria for positive definiteness of homogeneous polynomials with multiple variables (necessary due to custodial symmetry breaking effects from loops), we numerically analyze these constraints. Our results reveal that the parameter ranges allowed by positive definiteness differ significantly from those derived from tree-level BFB conditions. Notably, some regions previously excluded by tree-level BFB constraints remain viable under the one-loop RG-improved scalar potential. Besides, certain parameter spaces that satisfy tree-level BFB constraints with electroweak (EW) scale couplings should be excluded due to violations of positive definiteness in large field-value regions. Our numerical analysis, based on the new criteria for positive definiteness of homogeneous polynomials with multiple variables, not only revises the GM model's viability map but also provides a methodological template for stability studies in other extended Higgs models.

}

\begin{document}
\maketitle
\newpage
\section{Introduction}

The Standard Model (SM) of particle physics has been extensively tested by a multitude of experiments and is widely regarded as a successful low-energy effective theory at or below the electroweak (EW) scale. The discovery of the 125 GeV Higgs boson by the ATLAS and CMS collaborations ~\cite{ATLAS:higgs,CMS:higgs} at the Large Hadron Collider (LHC) marked the completion of the SM particle content. Although the observed Higgs boson properties align closely with SM predictions, the scalar sector responsible for electroweak symmetry breaking (EWSB) could still deviate significantly from the minimal SM framework. Indeed, an extended Higgs sector remains phenomenologically viable within current experimental bounds, potentially predicting additional scalar states beyond the 125 GeV resonance. Consequently, probing such new scalars and elucidating the origin of EWSB constitute primary objectives for the LHC and future colliders.

Extending the SM with additional scalars can be well-motivated theoretically.  For instance, the Georgi-Machacek (GM) model~\cite{GM,GM2},---which incorporates a complex $SU(2)_L$ triplet scalar $\chi$ with hypercharge $Y=1$ and a real $SU(2)_L$ triplet scalar $\xi$ with $Y=0$---preserves custodial symmetry nontrivially at tree level when the vacuum expectation values (VEVs) of the complex and real triplets align after EWSB, allowing large triplet VEVs of $\mathcal{O}(10)$ GeV. The $\chi$ triplet can be naturally accommodated into the Type-II neutrino seesaw mechanism to generate tiny neutrino masses~\cite{Triplet:neutrino1,Triplet:neutrino2,Triplet:neutrino3,Triplet:neutrino4,Triplet:neutrino5,Triplet:neutrino6,GM:neutrino,GM3}, though alternative neutrino mass-generation mechanisms are also feasible. Furthermore, the GM model can explain the baryon asymmetry of the universe (BAU) via electroweak baryogenesis, potentially producing detectable gravitational waves (GWs) during the electroweak phase transition (EWPT)~\cite{GM52,GM53,GM54,GM55}. Notably, the $W$-boson mass anomaly reported by the CDF II collaboration can be accommodated within the GM framework only when custodial-symmetry-breaking effects~\cite{GM48,GM49,GM50,GM51} are included.

Owing to its extended scalar sector, the GM model exhibits distinctive properties of phenomenological interest. It predicts that the SM-like Higgs couplings to $W$ and $Z$ gauge bosons can be significantly enhanced relative to SM expectations. In addition to the the three Goldstone modes absorbed by $W^\pm$ and $Z^0$, the model contains ten physical scalars, including doubly/singly charged and neutral scalars. Extensive studies have focused on searches for these new Higgs states at the LHC and future colliders (e.g., CEPC, ILC)~\cite{GM24,GM25,GM26,GM27,GM28,GM29,GM30,GM31,GM33,GM34,GM35,GM36,GM37,GM38,GM39,GM40,GM41,GM63}. 
The doubly charged Higgs boson within the custodial $SU(2)_V$ quintuplet exhibits rich phenomenology~\cite{GM16,GM17,GM18,GM19,GM20,GM21,GM22,GM23}, while the lighter CP-even neutral scalar (a custodial singlet) could correspond to the $\sim 95$ GeV resonance observed by CMS and ATLAS in diphoton searches~\cite{GM42,GM43}.

In GM model studies, theoretical constraints from perturbative unitarity and electroweak vacuum stability~\cite{GM57,GM58,GM59,GM60,GM61,GM62,GM64} must be satisfied alongside collider exclusion bounds~\cite{GM56}. The scalar potential's intricate structure permits complicated vacuum configurations, potentially allowing custodial-breaking minima to coexist with the desired electroweak symmetry breaking vacuum. For example, the vacuum
structure in the GM model with an exact $Z_2$ symmetry~\cite{GM60} may include charge-breaking minima, custodial-violating minima, or incorrect EWSB patterns. This imposes non-trivial constraints to ensure the custodial-preserving vacuum is the true global minimum. Critically, deeper vacua might emerge at large field values in GM model, analogous to the SM vacuum stability bound where the absence of such vacua sets a lower limit of 129 GeV for the Higgs mass to ensure absolute stability of EWSB vacuum up to the Planck scale~\cite{VS_SM1,VS_SM2}. However, prior GM studies adopted only necessary tree-level bounded-from-below~(BFB) conditions, which ignore quantum corrections that may dominantly reshape the potential's asymptotic behavior at large field values.

We propose to analyze the BFB bounds with the one-loop effective potential of GM model in this work.
It is fairly important to include the relevant radiative corrections for the GM model, not only for the BFB discussions, but also for imposing the experimental constraints. 
 Radiative corrections to the $\rho$ parameter at the one-loop level was discussed in various literatures. The one-loop radiative corrections to the $hZZ$ and $hWW$ couplings in GM model had been calculated in~\cite{Chiang:2017vvo} and they can be used to distinguish between the GM model and other custodial multi-Higgs models by the correlations between the deviations in the $hZZ$ and $hWW$ couplings from the standard model predictions. 
The renormalized vertices of the 125 GeV SM-like Higgs boson with the weak gauge bosons ($hVV$), fermions ($hFF$) and itself ($hhh$) in the Georgi-Machacek model at one-loop level had also been discussed in~\cite{Kikuchi:2013gba} and~\cite{Chiang:2018xpl}, rendering the possibility to distinguish between the GM model and other extended Higgs models by comparing the pattern of deviations in the Higgs boson couplings. The effects of custodial symmetry violation by loop effects and the high-energy behavior of the GM model had been discussed in~\cite{GM44} and~\cite{GM45}.

This work establishes positive definiteness constraints for the GM model's effective scalar potential to preclude deeper vacua in large field regimes, with emphasis on input parameter bounds. These constraints refine previous tree-level BFB/positivity conditions, enhancing our understanding of the vacuum structure and relaxing overly restrictive phenomenological limits. The paper is structured as follows: Section 2 reviews the GM model; Section 3 revisits existing constraints and introduces new analytical tools; Section 4 presents numerical results; and Section 5 summarizes conclusions.

\section{A brief review of the Georgi-Machacek model}

The Higgs sector of the GM model~\cite{GM} comprises the familiar $SU(2)_L$ complex doublet Higgs field, denoted as $({\phi}^+,~{\phi}^0)$ with hypercharge $Y=1/2$, and additional two $SU(2)_L$ triplet Higgs fields: a complex triplet Higgs denoted as $({\chi}^{++},~{\chi}^{+},~{\chi}^{0})$ with $Y=1$, and a real triplet denoted as $({\xi}^{+},~{\xi}^{0},~-{\xi}^{+*})$ with $Y=0$.
These fields can be expressed in $SU(2)_L \times SU(2)_R$-covariant matrix forms:
$$
\Phi = \begin{pmatrix} \phi^{0*} & \phi^+ \\ \phi^- & \phi^0 \end{pmatrix}, \quad
\Delta = \begin{pmatrix} \chi^{0*} & \xi^+ & \chi^{++} \\ \chi^- & \xi^0 & \chi^+ \\ \chi^{--} & \xi^- & \chi^{0} \end{pmatrix}~,
$$
where $(\phi^+)^* = -\phi^-$, $(\chi^+)^* = -\chi^-$, $(\chi^{++})^* = \chi^{--}$, and $(\xi^+)^* = -\xi^-$.  The transformations of $\Phi$ and $\Delta$ under the global $SU(2)_L \times SU(2)_R$ symmetry are given by
$$
\Phi \to U_L \Phi U_R^\dagger, \quad \Delta \to U_L \Delta U_R^\dagger~,
$$
with $U_{L,R} = \exp(i\theta_{L,R}^a T^a)$ and $T^a$ being the $SU(2)$ generators.

The most general gauge-invariant scalar potential that preserves the global $SU(2)_L \times SU(2)_R$ symmetry can be written as~\cite{GM58}
\beqa
V(\Phi,\Delta) &= & \frac{1}{2}m_{\Phi}^{2}\mathrm{tr}\left[\Phi^{\dagger}\Phi\right] + \frac{1}{2}m_{\Delta}^{2}\mathrm{tr}\left[\Delta^{\dagger}\Delta\right] + \lambda_{1}\left(\mathrm{tr}\left[\Phi^{\dagger}\Phi\right]\right)^{2} + \lambda_{2}\mathrm{tr}\left[\Phi^{\dagger}\Phi\right]\mathrm{tr}\left[\Delta^{\dagger}\Delta\right] \nn\\
& +& \lambda_{3}\mathrm{tr}\left[\left(\Delta^{\dagger}\Delta\right)^{2}\right] + \lambda_{4}\left(\mathrm{tr}\left[\Delta^{\dagger}\Delta\right]\right)^{2}
 - \lambda_{5}\mathrm{tr}\left[\Phi^{\dagger}\frac{\sigma^{a}}{2}\Phi\frac{\sigma^{b}}{2}\right]\mathrm{tr}\left[\Delta^{\dagger}T^{a}\Delta T^{b}\right] \nn\\
& -& M_{1}\mathrm{tr}\left[\Phi^{\dagger}\frac{\sigma^{a}}{2}\Phi\frac{\sigma^{b}}{2}\right](P^{\dagger}\Delta P)_{ab} - M_{2}\mathrm{tr}\left[\Delta^{\dagger}T^{a}\Delta T^{b}\right](P^{\dagger}\Delta P)_{ab},
\label{eq:GMpot}
\eeqa
where $\sigma^{a}$ are Pauli matrices, and $T^{a}$ are $SU(2)$ generators in the $3\times 3$ representation:
$$
T^{1}= \frac{1}{\sqrt{2}}\begin{pmatrix} 0 & 1 & 0 \\ 1 & 0 & 1 \\ 0 & 1 & 0 \end{pmatrix},\;
T^{2}=\frac{1}{\sqrt{2}}\begin{pmatrix} 0 & -i & 0 \\ i & 0 & -i \\ 0 & i & 0 \end{pmatrix},\;
T^{3}=\begin{pmatrix} 1 & 0 & 0 \\ 0 & 0 & 0 \\ 0 & 0 & -1 \end{pmatrix}.
$$
The matrix $P$ rotates triplets to the Cartesian basis:
$$
P = \frac{1}{\sqrt{2}}\begin{pmatrix} -1 & i & 0 \\ 0 & 0 & \sqrt{2} \\ 1 & i & 0 \end{pmatrix}.
$$
 It should be noted that the $M_1$ and $M_2$ terms vanish if a $Z_2$ symmetry ($\Delta \leftrightarrow -\Delta$) is imposed.

\subsection{Symmetry Breaking and Physical Scalars}
Neutral scalar fields decompose as:
$$
\phi^{0} \to \frac{v_{\phi}}{\sqrt{2}} + \frac{h_{\phi} + i a_{\phi}}{\sqrt{2}},\quad
\chi^{0} \to v_{\chi} + \frac{h_{\chi} + i a_{\chi}}{\sqrt{2}},\quad
\xi^{0} \to v_{\xi} + h_{\xi},
$$
where $v_{\phi}$, $v_{\chi}$, and $v_{\xi}$ are the corresponding VEVs. These VEVs trigger $SU(2)_L \times SU(2)_R$ breaking, with custodial $SU(2)_V$ symmetry preserved at tree level when $v_{\chi} = v_{\xi} \equiv v_{\Delta}$. Then the EWSB condition becomes:
$$
v^{2} = v_{\phi}^{2} + 4v_{\chi}^{2} + 4v_{\xi}^{2} = v_{\phi}^{2} + 8v_{\Delta}^{2} = \frac{1}{\sqrt{2}G_{F}} \approx (246~\mathrm{GeV})^{2}.
$$
The triplet contribution is quantified by:
$$
\sin\theta_H = \frac{2\sqrt{2}v_{\Delta}}{v}.
$$
After EWSB, three Goldstone modes are eaten by $W_\mu^\pm, Z_\mu^0$ and ten physical scalars remain:
\begin{itemize}
\item Two CP-even singlets ($h$, $H_1$)
\beqa
h&=& \cos\alpha \Re\phi^0-\sin\alpha \(\sqrt{\f{1}{3}}\xi^0+\sqrt{\f{2}{3}}\Re\chi^0~\)~,\nn\\
H&=& \sin\alpha \Re\phi^0+\cos\alpha \(\sqrt{\f{1}{3}}\xi^0+\sqrt{\f{2}{3}}\Re\chi^0~\)~.
\label{hH:combination}
\eeqa
\item One triplet $H_3$:
\beqa
H_3^+&=&-s_H \phi^++c_H\f{1}{\sqrt{2}}\(\chi^++\xi^+\)~,\nn\\
H_3^0&=&-s_H \Im\phi^{0}+ c_H\Im\chi^{0}~.
\label{H3:combination}
\eeqa
\item One quintuplet $H_5$
\beqa
H_5^{++}&=&\chi^{++}~,\nn\\
H_5^+&=&\f{1}{\sqrt{2}}\(\chi^+-\xi^+\)~,\nn\\
H_5^0&=&\sqrt{\f{2}{3}}\xi^0-\sqrt{\f{1}{3}}\Re\chi^0~.
\label{H5:combination}
\eeqa
\end{itemize}
which are classified by custodial $SU(2)_V$ representations. Tree-level custodial symmetry ensures tree-level mass degeneracy within each multiplet, with the corresponding masses given by
\beqa
m_5^2&=&\f{M_1}{4v_\chi}v_\phi^2+12 M_2 v_\chi+\f{3}{2}\la_5 v_\phi^2+8\la_3 v_\chi^2~,\nn\\
m_3^2&=&\f{M_1}{4v_\chi}\(v_\phi^2+8v_\chi^2\)+\f{\la_5}{2} \(v_\phi^2+8v_\chi^2\)~,\nn\\
m^2_{h,H_1}&=&\f{1}{2}\[{\cal{M}}_{11}^2+{\cal{M}}_{22}^2\mp\sqrt{\({\cal{M}}_{11}^2-{\cal{M}}_{22}^2\)^2+4\({\cal{M}}_{12}^2\)^2}\].
\label{multiplet:masses}
\eeqa 
with
\beqa
{\cal{M}}_{11}&=& 8\la_1 v_\phi^2~,\nn\\
{\cal{M}}_{12}&=&\f{\sqrt{3}}{2}v_\phi\[-M_1+4(2\la_2-\la_5)v_\chi\]~,\nn\\
{\cal{M}}_{22}&=&\f{M_1}{4v_\chi}v_\phi^2-6M_2 v_\chi+8(\la_3+3\la_4)v_\chi^2~.
\eeqa

 While perturbative unitarity, vacuum stability and indirect constraints from oblique parameters and the $Z$-pole observables, etc., allow $v_{\Delta} \leq 80~\mathrm{GeV}$, collider bounds on $H_{1}$, $H_{3}^{0}$, and $H_{5}^{0}$~\cite{GM32} restrict $v_{\Delta} \leq 40~\mathrm{GeV}$ ($\sin\theta_H < 0.45$).

\section{Positive definiteness constraints of effective scalar potential}
\label{sec:posdef}

\subsection{Theoretical Constraints}
\label{subsec:theoconstraints}

The parameters of the GM model must satisfy both experimental bounds and theoretical constraints, including perturbative unitarity bounds and vacuum stability constraints. The latter encompasses BFB constraints and requirements to avoid alternative minima, ensuring the electroweak-breaking custodial-preserving vacuum is the true global minimum.

Perturbative unitarity bounds arise from $S$-wave amplitudes for scalar boson elastic scattering~\cite{GM58,Aoki:2007ah} and require:
\begin{align}
&\sqrt{(6\lambda_1 - 7\lambda_3 - 11\lambda_4)^2 + 36\lambda_2^2} + |6\lambda_1 + 7\lambda_3 + 11\lambda_4| < 4\pi, \label{eq:unitarity1} \\
&\sqrt{(2\lambda_1 + \lambda_3 - 2\lambda_4)^2 + \lambda_5^2} + |2\lambda_1 - \lambda_3 + 2\lambda_4| < 4\pi, \label{eq:unitarity2} \\
&|2\lambda_3 + \lambda_4| < \pi,  \\
&|\lambda_2 - \lambda_5| < 2\pi~, \label{eqn:pertUnitarity}
\end{align}
for the quartic coupling parameters in the scalar potential.

Current BFB constraints require positive definiteness of tree-level scalar potential at large field values~\cite{GM58}. General analysis of necessary and sufficient tree-level BFB conditions for models containing $SU(2)_L$ triplets have been established in~\cite{Moultaka:2020dmb}. Given that quartic terms dominate the tree-level scalar potential at large field values, the quadratic and cubic terms in Eq.~(\ref{eq:GMpot}) become negligible when imposing BFB constraints. The quartic terms in the scalar potential are
\begin{equation}
V^{(4)}(r,\tan\gamma,\zeta,\omega) = \left[\lambda_1 + (\lambda_2 - \omega\lambda_5)\tan^{2}\gamma + (\zeta\lambda_{3} + \lambda_{4})\tan^{4}\gamma\right] \frac{r^{4}}{(1 + \tan^{2}\gamma)^{2}}, \label{eq:V4}
\end{equation}
where
\begin{align}
r &\equiv \sqrt{\mathrm{Tr}(\Phi^{\dagger}\Phi) + \mathrm{Tr}(\Delta^{\dagger}\Delta)}, \\
r^{2}\cos^{2}\gamma &\equiv \mathrm{Tr}(\Phi^{\dagger}\Phi), \\
r^{2}\sin^{2}\gamma &\equiv \mathrm{Tr}(\Delta^{\dagger}\Delta), \\
\zeta &\equiv \frac{\mathrm{Tr}(\Delta^{\dagger}\Delta\Delta^{\dagger}\Delta)}{|\mathrm{Tr}(\Delta^{\dagger}\Delta)|^{2}}, \\
\omega &\equiv \frac{\mathrm{Tr}(\Phi^{\dagger}\frac{\sigma^{a}}{2}\Phi\frac{\sigma^{b}}{2})\mathrm{Tr}(\Delta^{\dagger}T^{a}\Delta T^{b})}{\mathrm{Tr}(\Phi^{\dagger}\Phi)\mathrm{Tr}(\Delta^{\dagger}\Delta)}.
\end{align}
Parameters range as~\cite{GM58}:
\begin{equation}
r \in [0,\infty), \quad \gamma \in \left[0,\frac{\pi}{2}\right], \quad \zeta \in \left[\frac{1}{3},1\right], \quad \omega \in \left[-\frac{1}{4},\frac{1}{2}\right]. \label{eq:param_ranges}
\end{equation}

The BFB conditions of the tree-level scalar potential
give rise to the constraints on the quartic couplings
\begin{align}
\lambda_4 & > \begin{cases}
      -\frac{1}{3}\lambda_3 & \text{for } \lambda_3 \geq 0 \\
      -\lambda_3 & \text{for } \lambda_3 < 0
   \end{cases},  \\
\lambda_2 & > \begin{cases}
      \frac{1}{2}\lambda_5 - 2\sqrt{\lambda_1(\frac{1}{3}\lambda_3 + \lambda_4)} & \text{for } \lambda_5 \geq 0, \lambda_3 \geq 0 \\
      \omega_{+}(\zeta)\lambda_5 - 2\sqrt{\lambda_1(\zeta\lambda_3 + \lambda_4)} & \text{for } \lambda_5 \geq 0, \lambda_3 < 0 \\
      \omega_{-}(\zeta)\lambda_5 - 2\sqrt{\lambda_1(\zeta\lambda_3 + \lambda_4)} & \text{for } \lambda_5 < 0
   \end{cases},  \\
\lambda_1 & > 0, \label{eq:bfbcond2}
\end{align}
with $\omega_{\pm}(\zeta)$ given by:
\begin{equation}
\omega_{\pm}(\zeta) = \frac{1}{6}\left(1 - \sqrt{\frac{3}{2}\left(\zeta - \frac{1}{3}\right)}\right) \pm \frac{\sqrt{2}}{3} \left[\left(1 - \sqrt{\frac{3}{2}\left(\zeta - \frac{1}{3}\right)}\right)\left(\frac{1}{2} + \sqrt{\frac{3}{2}\left(\zeta - \frac{1}{3}\right)}\right)\right]^{1/2}, \label{eq:omega_func}
\end{equation}
which must hold for all $\zeta$. Crucially, this formulation can be applied only when
 the potential can be rewritten as a quadratic form, which fails when quantum corrections introduce custodial symmetry breaking via $U(1)_Y$ hypercharge interactions.

\subsection{Effective Potential in the GM Model}
\label{subsec:effectivepot}

Standard BFB constraints require positive definiteness of tree-level scalar potential, but quantum corrections can reshape the potential, potentially violating positive definiteness and creating deeper vacua at large field values. This may render the custodial-symmetry preserving EWSB vacuum unstable, leading to unrealistic phenomenology.

To determine the vacuum structures and the locations of the minima, we need to know the effective potential of the GM model, which is the sum of all one-particle irreducible diagrams with zero external momenta. At one-loop order, the effective potential can be calculated from the general form
\begin{equation}
V_{\text{eff}}^{1} = V_{0} + V_{\text{CW}}(\mu_{R}) = V_{0} + \sum_{i} n_{i} \frac{s_{i}m_{i}^{4}}{64\pi^{2}} \left( \log\left( \frac{m_{i}^{2}}{\mu_{R}^{2}} \right) - c_{i} \right), \label{eq:Veff}
\end{equation}
for each individual mode $i$ with multiplicity $n_i$, where $V_0$ is the classical potential, $s_i = -1$ for fermions and $+1$ for bosons. Here $V_0$ takes the same form as Eq.(\ref{eq:GMpot}) in bare parameters and $V_{CW}$ is the zero temperature one-loop effective potential renormalized at the scale $\mu_R$~\cite{cw}. It was noted in~\cite{sher} that, although the one-loop effective potential can be expressed in a simple and compact way, it is valid only if $\al \log(\phi_i/\phi_j)\ll 1$, where $\al$ is the largest couplings in the model and $\phi_a$($\phi_b$) is the largest (smallest) value considered. Summing over potentially large logarithms, one can arrive at the renormralization group-improved (RG-improved) effective potential, which is valid for all field values as long as the couplings are small. In fact, the $L$-loop effective potential improved by $(L+1)$-loop
 renormalization group equations (RGEs) resume all $L$-th to leading logarithm contributions~\cite{Kastening}.

In practice, the RG-improved effective scalar potential can be obtained by solving the RGEs.
The effective potential is not affected by the change of the renormalization parameters, whose effects can be absorbed in the changes of the couplings and fields. Therefore, it satisfies
\begin{equation}
\left[ \frac{\partial}{\partial \ln \mu} + \sum_{g_i} \beta_{g_i} \frac{\partial}{\partial g_i} - \gamma_m m \frac{\partial}{\partial m} - \sum_{\phi_j} \gamma_j \right] V = 0, \label{eq:RGE}
\end{equation}
where beta functions and anomalous dimensions are:
\begin{align}
\beta_{g_i} &= \frac{\partial g_i}{\partial \ln \mu}, \\
\gamma_m &= -\frac{\partial \ln m}{\partial \ln \mu} = \frac{\partial \ln Z_m}{\partial \ln \mu}, \\
\gamma_j &= \frac{1}{2} \frac{\partial \ln Z_{\phi_j}}{\partial \ln \mu}. \label{eq:anomalous}
\end{align}

The RG-improved effective scalar potential is given by
\beqa
V(\phi_i)=\f{1}{2}m_i^2(t)G_{\phi_i}^2(t)\phi_i^2+\f{1}{4}\la_i(t)G_{\phi_i}^4(t)\phi_i^4~,
\eeqa
with $t=\ln\f{\phi_i}{\mu}$ and
\beqa
G_{\phi_i}(t)=\exp{\int\limits_{0}^t\ga_j(\mu) d (\ln \mu)}.
\eeqa
The standard one-loop effective potential can be derived easily from this result with the approximation of constant beta function $\beta_{\lambda}$ and vanishing anomalous dimension $\ga_j$.

In the large field value regions $(\Delta \gg v_{\Delta}, \Phi \gg v_{\phi})$, the effective scalar potential can be very well approximated by the RG-improved tree-level expression:
\begin{align}
V(\Phi,\Delta) =& \lambda_1(\mu) \left( \mathrm{tr} [\Phi^{\dagger}\Phi] \right)^2 + \lambda_2(\mu) \left( \mathrm{tr} [\Delta^{\dagger}\Delta] \right)^2 + \lambda_3(\mu) \mathrm{tr} \left[ (\Delta^{\dagger}\Delta)^2 \right] \nonumber \\
&+ \lambda_4(\mu) \mathrm{tr} [\Phi^{\dagger}\Phi] \mathrm{tr} [\Delta^{\dagger}\Delta] + \lambda_5(\mu) \mathrm{tr} \left[ \Phi^{\dagger}\frac{\sigma^{a}}{2}\Phi\frac{\sigma^{b}}{2} \right] \mathrm{tr} \left[ \Delta^{\dagger}T^{a}\Delta T^{b} \right] \nonumber \\
&+ V_{\text{c-v}}, \label{eq:Vlargefield}
\end{align}
where $V_{\text{c-v}}$ denotes custodial-symmetry-violating terms from quantum effects, and $\mu \sim \mathcal{O}(h_{\phi;0}, h_{\chi;0}, \cdots)$. Similar to the scalar potential in SM, whose quartic coupling $\lambda$ becomes negative around ${10}^{11}$ GeV, the quartic couplings $\la_i$ in GM model may also change signs at large field values, possibly spoiling the positive definiteness of effective scalar potential and leading to the emergence of new deeper vacuum at large field values. It is enough to determine whether deeper vacua emerge in various large field value regions with such RG-improved tree-level scalar potential.

The custodial symmetry preserved at tree level scalar potential is explicitly broken by $U(1)_Y$ gauge interactions at quantum level.  When the custodial symmetry is imposed at the electroweak scale, renormalization group evolution inevitably generates custodial-symmetry-breaking operators. 
Therefore, it is not consistent to adopt the Lagrangian with a custodial-symmetry-preserving scalar potential for proper renormalization group evolution description, as new custodial symmetry breaking terms will be generated so as that divergent loop corrections related to such terms can not be redefined properly. Consequently, one should begin with the most general $SU(3)_C \times SU(2)_L \times U(1)_Y$ gauge invariant scalar potential~\cite{GM44,GM45},  which includes explicitly all possible custodial symmetry violation terms that can be consistent with the gauge symmetry.

Detailed discussions on the generations of the custodial-symmetry-breaking operators at high energies can be found in~\cite{GM44}, assuming no custodial-symmetry-violation terms at the EW scale. They found that, in some parameter region, the custodial symmetry breaking effects can be well controlled up to the GUT scale. On the other hand, if a custodial symmetry is chosen to be preserved at high energy scale in some UV completion of GM model, the measured value of the eletroweak $\rho$ parameter (along with perturbative unitarity) can constrain stringently the scale
of the UV completion over almost all of the parameter
space~\cite{GM45}. The two approaches, used to explore the custodial symmetry breaking effects of the GM model induced by RGE, are equivalent.

The most general $SU(3)_C \times SU(2)_L \times U(1)_Y$ invariant GM model scalar potential with all possible custodial symmetry violation terms can be written as:
\begin{align}
V(\phi,\chi,\xi) =& m_{\phi}^{2}(\phi^{\dagger}\phi) + m_{\chi}^{2}\mathrm{tr}(\chi^{\dagger}\chi) + m_{\xi}^{2}\mathrm{tr}(\xi^{2}) \nonumber \\
& + \mu_1 \phi^{\dagger}\xi\phi + \mu_2 \left[ \phi^{T}(i\tau_2)\chi^{\dagger}\phi + \mathrm{h.c.} \right] + \mu_3 \mathrm{tr}(\chi^{\dagger}\chi\xi) \nonumber \\
& + \lambda (\phi^{\dagger}\phi)^2 + \rho_1 [\mathrm{tr}(\chi^{\dagger}\chi)]^{2} + \rho_2 \mathrm{tr}(\chi^{\dagger}\chi\chi^{\dagger}\chi) + \rho_3 \mathrm{tr}(\xi^{4}) \nonumber \\
& + \rho_4 \mathrm{tr}(\chi^{\dagger}\chi) \mathrm{tr}(\xi^{2}) + \rho_5 \mathrm{tr}(\chi^{\dagger}\xi) \mathrm{tr}(\xi\chi) + \sigma_1 \mathrm{tr}(\chi^{\dagger}\chi)\phi^{\dagger}\phi \nonumber \\
& + \sigma_2 \phi^{\dagger}\chi\chi^{\dagger}\phi + \sigma_3 \mathrm{tr}(\xi^{2})\phi^{\dagger}\phi + \sigma_4 \left( \phi^{\dagger}\chi\xi(i\tau_2)\phi^{*} + \mathrm{h.c.} \right), \label{eq:Vgeneral}
\end{align}
with field representations:
\begin{align*}
\phi = \begin{pmatrix} \phi^{+} \\ \phi^{0} \end{pmatrix}, \quad
\chi = \begin{pmatrix} \frac{\chi^{+}}{\sqrt{2}} & -\chi^{++} \\ \chi^{0} & -\frac{\chi^{+}}{\sqrt{2}} \end{pmatrix}, \quad
\xi = \begin{pmatrix} \frac{\xi^{0}}{\sqrt{2}} & -\xi^{+} \\ -\xi^{-} & -\frac{\xi^{0}}{\sqrt{2}} \end{pmatrix}.
\end{align*}

The custodial-symmetric limit corresponds to~\footnote{For simplicity, we assume $\mu_2$ and $\sigma_4$ take real values.}:
\begin{eqnarray}
m_\phi^2&=&2m_\Phi^2,~ m_\chi^2=2m_\Delta^2,~ m_\xi^2= m_\Delta^2,~
\mu_1=-\frac{M_1}{\sqrt{2}},~ \nonumber \\
\mu_2&=&-\frac{M_1}{2},~
\mu_3=6\sqrt{2}M_2,
\lambda=4\lambda_1,~
\rho_1=4\lambda_2+6\lambda_3,~\nonumber \\
\rho_2&=&-4\lambda_3,~
\rho_3=2(\lambda_2+\lambda_3),~
\rho_4=4\lambda_2,~
\rho_5=4\lambda_3,~\nonumber \\
\sigma_1 &=& 4\lambda_4-\lambda_5,~
 \sigma_2 =2\lambda_5,~
 \sigma_3 =2\lambda_4,~
 \sigma_4 =\sqrt{2}\lambda_5.
\label{rel}
\end{eqnarray}
 RGE evolution generates deviations from these relations, quantified the custodial symmetry
breaking effects by the loops involving $U(1)_Y$ gauge interactions.

To proceed with the RGE analysis, we need to derive the $\beta$-functions for the scalar quartic couplings $\lambda,~\rho_{1,2,3,4,5},~\sigma_{1,2,3,4}$, the gauge couplings $g_i$, and the Yukawa couplings $Y_{t,b,\tau}$ etc. Two-loop beta-functions for a generic quantum field theory can be found in~\cite{Machacek:1983tz, Machacek:1983fi, Machacek:1984zw}.
For the GM model, the one-loop RGEs has already been given in~\cite{GM44,GM45}, and the two-loop RGEs can be found in~\cite{GM64}. In realistic numerical studies, the Mathematica package SARAH \cite{SARAH1,SARAH2,SARAH3,SARAH4,SARAH5} can be used for RGE numerical studies and generate the source codes for the spectrum-generator package SPheno \cite{SPheno1,SPheno2}. We list the expressions of relevant $\beta$ functions in Appendix \ref{app:A}.

\subsection{Positive Definiteness Analysis}
\label{subsec:posdefanalysis}
To understand the vacuum structure of the GM model, it is of crucial importance to analyze the positive definiteness constraints of the one-loop RG-improved tree-level scalar potential. Such positive definiteness constraints can ensure that the custodial $SU(2)$ symmetry preserving EWSB minimum is
the true global minimum and no deeper vacua emerge in various large field value regions.

Defining the following auxiliary parameters:
\begin{align}
\zeta &\equiv \frac{\mathrm{tr}(\chi^{\dagger}\chi\chi^{\dagger}\chi)}{[\mathrm{tr}(\chi^{\dagger}\chi)]^{2}}, \quad
\omega \equiv \frac{\mathrm{tr}(\chi^{\dagger}\xi)\mathrm{tr}(\xi\chi)}{\mathrm{tr}(\chi^{\dagger}\chi)\mathrm{tr}(\xi^{2})}, \quad
\eta \equiv \frac{\phi^{\dagger}\chi\chi^{\dagger}\phi}{\mathrm{tr}(\chi^{\dagger}\chi)\phi^{\dagger}\phi}, \nonumber \\
\kappa &\equiv \frac{[\phi^{\dagger}\chi\xi\tilde{\phi} + \mathrm{h.c.}]}{\sqrt{\mathrm{tr}(\phi^{\dagger}\phi)\mathrm{tr}(\tilde{\phi}^{\dagger}\tilde{\phi})\mathrm{tr}(\chi^{\dagger}\chi)\mathrm{tr}(\xi^{2})}}, \quad \tilde{\phi} \equiv i\tau_2\phi^{*}, \label{eq:auxparams}
\end{align}
the ranges of the parameters satisfy
\beqa
\zeta\in [\f{1}{2},1]~,~\omega\in [0,1]~,~\eta\in [0,1]~,~~\ka\in [-1,1],
\eeqa
by Cauchy-Schwarz inequality.

Neglecting the mass terms, the potential becomes a homogeneous quartic polynomial:
\beqa
V(z_1,z_2,z_3) &=& \lambda z_1^{4} + (\rho_1 + \zeta\rho_2) z_2^{4} + \rho_3 z_3^{4} + (\rho_4 + \omega\rho_5) z_2^{2}z_3^{2}\nn\\
 &+& (\sigma_1 + \eta\sigma_2) z_1^{2}z_2^{2} + \sigma_3 z_1^{2}z_3^{2} + \kappa\sigma_4 z_1^{2}z_2 z_3, \label{eq:Vhomogeneous}
\eeqa
where
\beqa
z_1^2 = \phi^{\dagger}\phi, ~z_2^2 = \mathrm{tr}(\chi^{\dagger}\chi), ~z_3^2 = \mathrm{tr}(\xi^{2})~.
\eeqa

When the scalar potential is not positive definite in large field value regions, the emergence of negative potential energy in certain directions indicates that vacua deeper than the custodial-preserving EWSB vacuum can develop in those regions. Given that the potential energy of the EWSB vacuum is conventionally normalized to approximately zero, the positive definiteness requirement for $V(z_1, z_2, z_3)$ serves as a necessary criterion to ensure the stability of the custodial-symmetry preserving EWSB vacuum configuration, thereby preventing quantum tunneling transitions into deeper vacua at asymptotically large field values.

Unlike the BFB condition applicable to the tree-level scalar potential in Eq.~(\ref{eq:GMpot}) (after neglecting the subleading dimensional terms), the scalar potential $V(z_1, z_2, z_3)$ cannot be reformulated as a quadratic form because custodial symmetry-breaking terms are dynamically generated during renormalization group evolution when quantum effects from $U(1)_Y$ gauge interactions are included. Consequently, the conventional Sylvester's criterion for positive definiteness becomes inapplicable to $V(z_1, z_2, z_3)$. This necessitates the development of new positive-definiteness criteria specifically designed for homogeneous polynomials of even degree $d > 2$ with multiple variables, which form the methodological foundation of our analysis.

\subsection{\label{criteria}New Positive Definiteness Criterion}

 General discussions from algebraic geometry~\cite{GKZ,Fernando} indicate that  positive definite of a homogeneous polynomial $F$ of even degree $d$ with multiple $n$ variables requires the characteristic polynomial of $F$
\beqa
\chi(F,J)(t)\equiv\Delta(F+tJ),
\eeqa
with $\Delta(F)$ the discriminant of $F$, to satisfy
\beqa
\chi(F,J)(t)>0~, ~~~\forall t\geq 0~.
\eeqa
for
\beqa
J(x)=\sum\limits_{1\leq j\leq n} x_j^d~.
\eeqa
As the explicit expression of resultant  $Res(F_{x_1},F_{x_2},\cdots,F_{x_n})$ are needed for the expressions of $\Delta(F)$, such a criterion is not useful for realistic numerical studies.

Since custodial breaking prevents reduction to a quadratic form, we employ a numerical method based on Lyapunov functions~\cite{Miao,Miao1,Bose,Bose1} to determine if a homogeneous polynomial with multiple variables is positive definite, which is correct in the sense of probability. For a homogeneous polynomial $P(\mathbf{x})$ of degree $m$ with multiple variables
\beqa
 P(x)=\sum\limits_{i_1+i_2+\cdots+i_n=m}\sum a_{i_1 \cdots i_n}x_1^{i_1}\cdots x_n^{i_n},
\eeqa
the following propositions hold:
\begin{itemize}
\item $P(\mathbf{x})$ is positive definite iff positive on the sphere $S_r = \{\mathbf{x} | \|\mathbf{x}\| = r > 0\}$ (with $||x||=\sum\limits_{j=1}^n x_{j}^2$).

\item $P(\mathbf{x})$ is positive definite iff
    \beqa
    \inf\limits_{||x||=1} P(\mathbf{x})>0~.
    \eeqa
\item  Let $V\equiv \{x| P(x)=c, x\in S_r, c~{\rm is~a~constant} \}$. If $V$ is connected, it can be considered as a point. Then $P(x) $ can only have finite local minimum points on $S_r$.

\item  Consider partitioning $S_r$ into $k$ regions $(\{ D_1, \dots, D_k \})$ where:
    \begin{itemize}
        \item Each region $D_i$ contains exactly one local minimum
        \item Every point in $D_i$ has a descent curve to its local minimum
        \item $S_i$ denotes the hyperarea (measure) of $D_i$.
    \end{itemize}

Probability of missing the global minimum after $K$ trials is:
\begin{equation}
p_K = 1 - \frac{C_K^k k! \prod_{i=1}^k (S_i / S_0) (K - k)^k}{K^k}, \label{eq:prob}
\end{equation}
where $S_i$ are areas of partition regions on $S_r$. Obviously, $\lim\limits_{K\ra\infty}p_K=0$.

\end{itemize}
The probability of correctness satisfies:
\begin{equation}
p \geq 1 - k \sum_{i=1}^k \left(1 - \frac{S_i}{S_0}\right)^K. \label{eq:prob_bound}
\end{equation}
We use $K=100$ randomly varying initial points to determine global minima in our numerical studies.

The computational procedure to find the minimum is
\begin{enumerate}
    \item \textbf{Local minimum search}: From any starting point on the unit hypersphere $S_r$, local minima of $P(x)$ can be found by gradient descent optimization algorithms with a continuous descent curve.

    \item \textbf{Random sampling}: By randomly varying initial points $x_0$, we obtain a sequence of local minima. As the number of random starting points increases, the probability of discovering \textbf{all local minima approaches 1}.

\end{enumerate}

\subsection{Vacuum Structure Analysis}
As we are interested in the vacuum structure of the GM model, the positive definiteness bounds can be derived with the VEVs of the relevant fields. The 6 symmetry transformations in $SU(2)_L\tm SU(2)_R$ can be used to eliminate the redundant VEVs~\cite{GM60}. We can choose to eliminate three of the four components of the doublet by the three broken symmetry transformation, leaving only $v_\phi$. Besides, the three custodial symmetry transformations can be used to eliminate both the real and imaginary components of $\chi^{++}$ and the imaginary component of $\xi^+$. Assuming CP conservation, the VEV of the imaginary components of $\chi^+$ and $\chi^0$ should vanish, that is, $v_{\chi;+}^I=v_{\chi;0}^I=0$. Therefore, the vacuum configuration takes the following form:
\begin{align}
\langle \phi \rangle &= \begin{pmatrix} 0 \\ \frac{v_{\phi}}{\sqrt{2}} \end{pmatrix}, \quad
\langle \chi \rangle = \begin{pmatrix} \frac{v_{\chi;+}^{R}}{\sqrt{2}} & 0 \\ v_{\chi;0}^{R} & -\frac{v_{\chi;+}^{R}}{\sqrt{2}} \end{pmatrix}, \quad
\langle \xi \rangle = \begin{pmatrix} \frac{v_{\xi;0}}{\sqrt{2}} & -v_{\xi;+} \\ -v_{\xi;+} & -\frac{v_{\xi;0}}{\sqrt{2}} \end{pmatrix}. \label{eq:vevs}
\end{align}

The RG-improved potential, after neglecting the dimensional terms, equals to:
\begin{align}
V =& \frac{\lambda}{4}v_{\phi}^{4} + \rho_1 \left( v_{\chi;+}^{2} + v_{\chi;0}^{2} \right)^2 + \rho_2 \left( \frac{1}{2}v_{\chi;+}^{4} + v_{\chi;0}^{4} + 3v_{\chi;+}^{2}v_{\chi;0}^{2} \right) \nonumber \\
& + 2\rho_3 \left( \frac{1}{4}v_{\xi;0}^{4} + v_{\xi;+}^{4} + v_{\xi;0}^{2}v_{\xi;+}^{2} \right) + \rho_4 \left( v_{\chi;+}^{2} + v_{\chi;0}^{2} \right) \left( 2v_{\xi;+}^{2} + v_{\xi;0}^{2} \right) \nonumber \\
& + \rho_5 \left( v_{\xi;0}^{2}v_{\chi;+}^{2} + v_{\xi;+}^{2}v_{\chi;0}^{2} - 2v_{\xi;0}v_{\chi;+}v_{\xi;+}v_{\chi;0} \right) \nonumber \\
& + \frac{\sigma_1}{2}v_{\phi}^{2} \left( v_{\chi;+}^{2} + v_{\chi;0}^{2} \right) + \frac{\sigma_2}{4}v_{\phi}^{2}v_{\chi;+}^{2} + \frac{\sigma_3}{2} \left( v_{\xi;0}^{2} + 2v_{\xi;+}^{2} \right) v_{\phi}^{2} \nonumber \\
& + \frac{\sigma_4}{2\sqrt{2}} v_{\phi}^{2} \left( v_{\chi;+}v_{\xi;+} + v_{\chi;0}v_{\xi;0} \right), \label{eq:Vrg}
\end{align}
in the large field value regions with the quartic couplings evaluated at RG scale $\mu$. Here we denote $v_{\chi;+}^R\equiv v_{\chi;+}$ and $v_{\chi;0}^R\equiv v_{\chi;0}$, respectively. Although the value of the scalar potential should be evaluated at the stationary point for which these VEVs hold, the minimization conditions for the VEVs can constrain only the dimensional parameters, which plays negligible roles in the derivation of the positive definiteness bounds in the large field value regions where the quartic terms dominate.

To ensure vacuum stability, it becomes imperative that no lower-lying vacua--other than the conventional custodial-preserving vacuum--develop in regions of large field values. This stability criterion translates to the requirement that the scalar potential $V(v_\phi, v_\chi, v_\xi)$ must remain positive definite along all asymptotic directions in field space, similar to the setting of ordinary vacuum stability bounds in the Standard Model. The form of the scalar potential is a homogeneous polynomial with multiple variables, whose positive definiteness can be determined by the methods in subsection~\ref{criteria}.

When $\rho_5, \sigma_4 \approx 0$, the potential simplifies to a quadratic form:
\begin{equation}
V = \begin{pmatrix} v_{\phi}^2 & v_{\chi;+}^2 & v_{\chi;0}^2 & v_{\xi;+}^2 & v_{\xi;0}^2 \end{pmatrix}
\left(\bea{ccccc}\f{\la}{4}&\f{2\sigma_1+\sigma_2}{8}&\f{\sigma_1}{4}&\f{\sigma_3}{2}&\f{\sigma_3}{4}\\
\f{2\sigma_1+\sigma_2}{8}&\f{2\rho_1+\rho_2}{2}&\f{3\rho_2}{2}&\rho_4&\f{\rho_4}{2}\\
\f{\sigma_1}{4}&\f{3\rho_2}{2}&\rho_1+\rho_2&\rho_4&\f{\rho_4}{2}\\
\f{\sigma_3}{4}&\rho_4&{\rho_4}&2\rho_3&\rho_3\\
\f{\sigma_3}{2}&\f{\rho_4}{2}&\f{\rho_4}{2}&\rho_3&\f{\rho_3}{2}\eea\right)
\begin{pmatrix} v_{\phi}^2 \\ v_{\chi;+}^2 \\ v_{\chi;0}^2 \\ v_{\xi;+}^2 \\ v_{\xi;0}^2 \end{pmatrix}. \label{eq:Vquadratic}
\end{equation}
 Positive definiteness requires all eigenvalues are positive. By Sylvester's criterion, such a condition is equivalent to the requirements that all principal minors are positive:
\beqa
&\lambda>0~,
~~~\det\left(\bea{cc}\f{\la}{4}&\f{2\sigma_1+\sigma_2}{8}\\\f{2\sigma_1+\sigma_2}{8}&\f{2\rho_1+\rho_2}{2}\eea\right)>0~,~~~\det\left(\bea{ccc}\f{\la}{4}&\f{2\sigma_1+\sigma_2}{8}&\f{\sigma_1}{4}\\\f{2\sigma_1+\sigma_2}{8}&\f{2\rho_1+\rho_2}{2}&\f{3\rho_2}{2}\\\f{\sigma_1}{4}&\f{3\rho_2}{2}&\rho_1+\rho_2\eea\right)>0~,\nn\\
&\det\left(\bea{cccc}\f{\la}{4}&\f{2\sigma_1+\sigma_2}{8}&\f{\sigma_1}{4}&\f{\sigma_3}{2}\\\f{2\sigma_1+\sigma_2}{8}&\f{2\rho_1+\rho_2}{2}&\f{3\rho_2}{2}&\rho_4\\\f{\sigma_1}{4}&\f{3\rho_2}{2}&\rho_1+\rho_2&\rho_4\\\f{\sigma_3}{4}&\rho_4&{\rho_4}&2\rho_3\eea\right)>0~,~~~\det\left(\bea{ccccc}\f{\la}{4}&\f{2\sigma_1+\sigma_2}{8}&\f{\sigma_1}{4}&\f{\sigma_3}{2}&\f{\sigma_3}{4}\\\f{2\sigma_1+\sigma_2}{8}&\f{2\rho_1+\rho_2}{2}&\f{3\rho_2}{2}&\rho_4&\f{\rho_4}{2}\\\f{\sigma_1}{4}&\f{3\rho_2}{2}&\rho_1+\rho_2&\rho_4&\f{\rho_4}{2}\\\f{\sigma_3}{4}&\rho_4&{\rho_4}&2\rho_3&\rho_3\\\f{\sigma_3}{2}&\f{\rho_4}{2}&\f{\rho_4}{2}&\rho_3&\f{\rho_3}{2}\eea\right)>0~.\nn\\
\eeqa

\section{Numerical Results}
\label{sec:numresults}

\subsection{Positive Definiteness Constraints Without Collider Bounds}
\label{subsec:noCollider}

Given the forms of the RG-improved tree-level scalar potential (\ref{eq:Vrg}) at each energy scale, we can assess whether the potential remains positive definite up to the Planck scale or the Landau pole scale with the methods in subsection~\ref{criteria}. This ensures that no new, deeper vacua emerge in the large field value regions, thereby guaranteeing the stability of the ordinary EWSB vacuum that preserves the custodial symmetry.

We initiate our numerical investigation by examining positive definiteness constraints in the absence of collider limits--a scenario that allows us to isolate the intrinsic theoretical constraints of the model. The set of input parameters is chosen to satisfy the theoretical perturbative unitarity bound. The input parameters $\lambda_{1,2,3,4,5}$ and $\sin\theta_H$ span theoretically motivated ranges:
\begin{equation}
\lambda_{1,2,3,4,5} \in (-\sqrt{4\pi}, \sqrt{4\pi}), \quad \sin\theta_H \in (0, 0.45).
\end{equation}
 The dimensional parameters $m^2_{\phi,\chi,\xi}$ are set to zero because their contribution becomes negligible in the large field value regions where the quartic terms dominate. This choice deliberately focuses our analysis on the asymptotic behavior of the potential.

 To determine the scale dependence of the quartic couplings in the RG-improved potential, we evolve the input parameters via renormalization group equations from the electroweak scale ($M_Z = 91.2~\text{GeV}$) to the Planck scale ($M_{\text{Planck}} \approx 1.12 \times 10^{19}~\text{GeV}$) or to the Landau pole scale (defined as the scale where any coupling diverges). The gauge coupling values at the electroweak scale are adopted from Ref.~\cite{Antusch:2013jca}. The boundary values for the scalar quartic couplings at the electroweak scale are chosen according to Eq.~(\ref{rel}) to ensure that the vacuum configuration preserves custodial symmetry.

Knowing the forms of the RG-improved tree-level scalar potential at each scale, we can verify that the potential remains strictly positive all the way up to the Planck scale (or to the Landau-pole scale), thereby guaranteeing that no deeper minima develop at large field values and that the standard EWSB vacuum, which preserves custodial symmetry, is absolutely stable.
 For each set of input parameters, it is necessary to verify that the conditions for positive definiteness hold both at the highest scale where the renormalization group equations are terminated (i.e., the Planck scale or the Landau pole scale) and at every scale during the RGE evolution upon 10 TeV. In practical numerical studies, it suffices to verify the positive definiteness of the scalar potential both at the local minima and at the saddle points, where one or more of the quartic couplings become negative.


To quantify custodial symmetry breaking induced by RGE evolution, we define deviation parameters $\delta_i$ at high scales:
\begin{align}
\delta_1 &\equiv \rho_3 - \rho_1/2 - \rho_2/4,
~~~~\delta_2 \equiv \rho_4 - \rho_1 - 3\rho_2/2, \nn \\
\delta_3 &\equiv \rho_5 + \rho_2,
~~~~~~~~~~~~~~~~\delta_4  \equiv \sigma_3 - \sigma_1/2 - \sigma_2/4, \nn\\
\delta_5 &\equiv \sigma_4 - \sigma_2/\sqrt{2}.
\end{align}

Our detailed numerical analysis demonstrates that when custodial breaking is small, the traditional BFB constraints~(\ref{eq:bfbcond2}) for $\lambda_i(\mu)$ align with our positive definiteness bounds. However, in regions where $\delta_i$ become significant, naively applying~(\ref{eq:bfbcond2}) by identifying the running couplings through the tree-level relations:
\beqa
\lambda_1 = \frac{\lambda}{4},\quad \lambda_2 = \frac{\rho_4}{4},\quad \lambda_3 = \frac{\rho_5}{4},\quad \lambda_4 = \frac{\sigma_3}{2},\quad \lambda_5 = \frac{\sqrt{2}\sigma_4}{2}~,
\eeqa
yields erroneous conclusions compared to our direct numerical evaluation of positivity. This demonstrates the necessity of our more comprehensive approach for proper treatment of custodial-breaking effects.


\begin{figure}[htbp]
\begin{center}
\includegraphics[width=12cm]{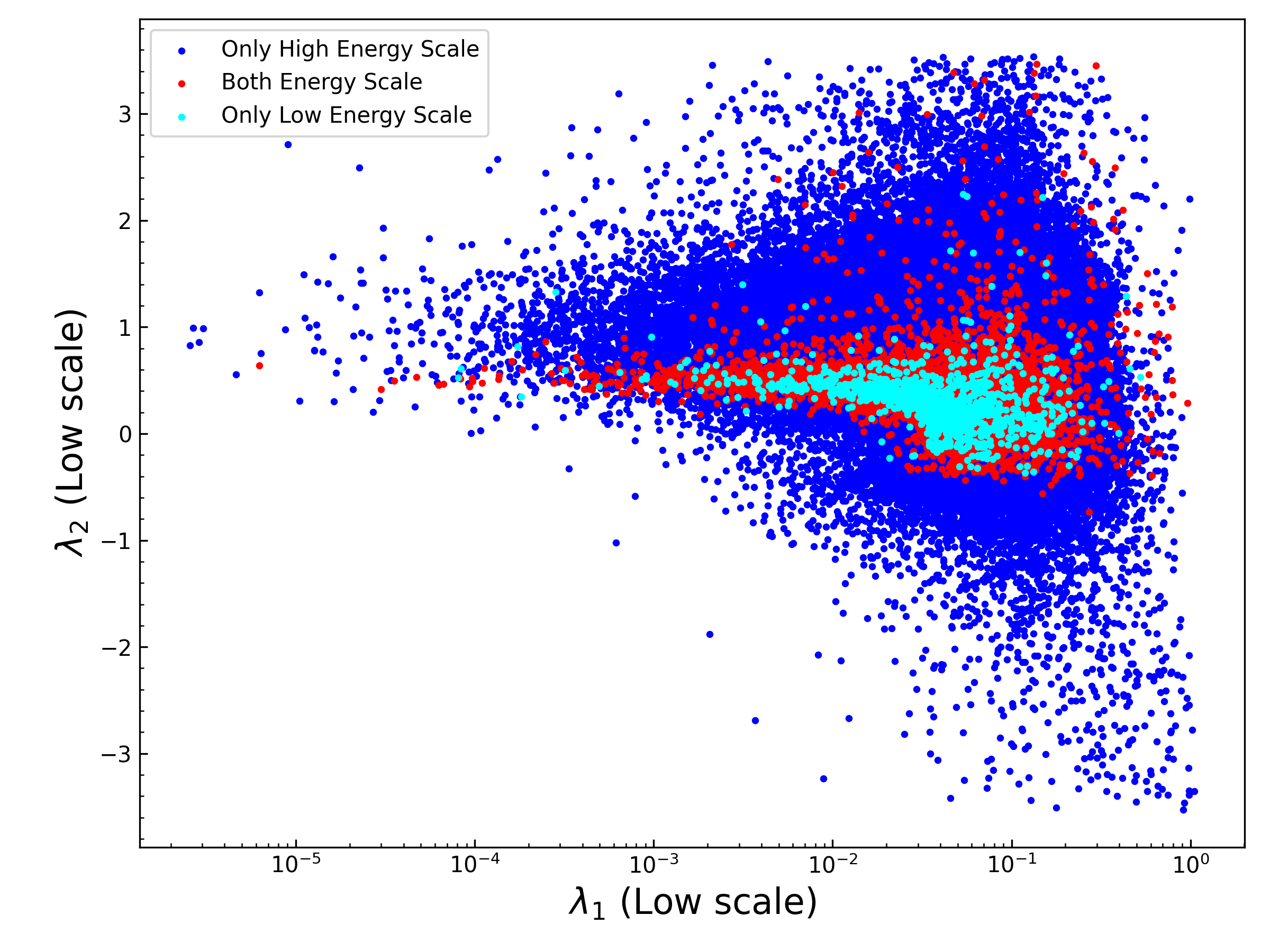}\\
\includegraphics[width=7.5cm]{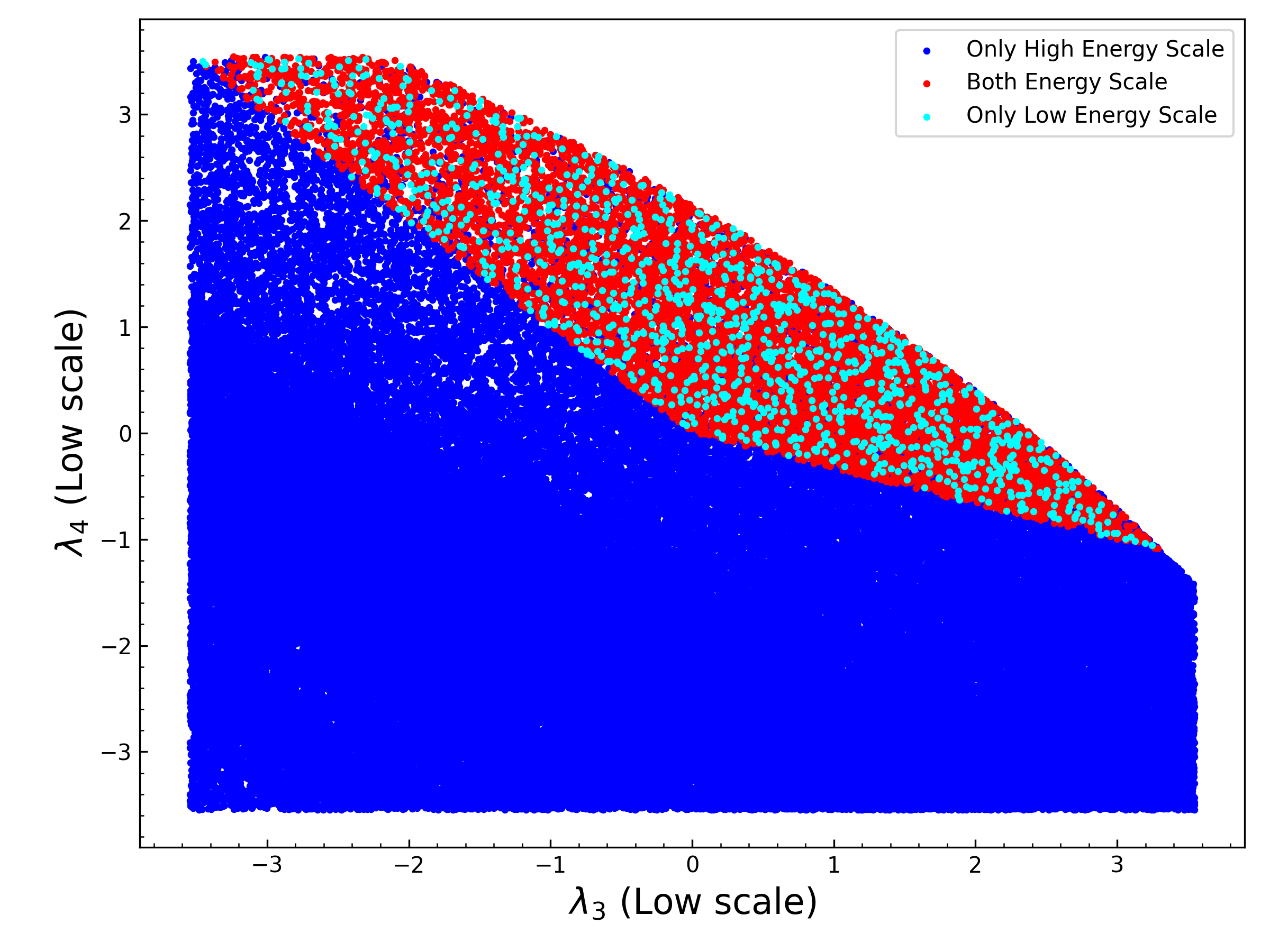}
\includegraphics[width=7.5cm]{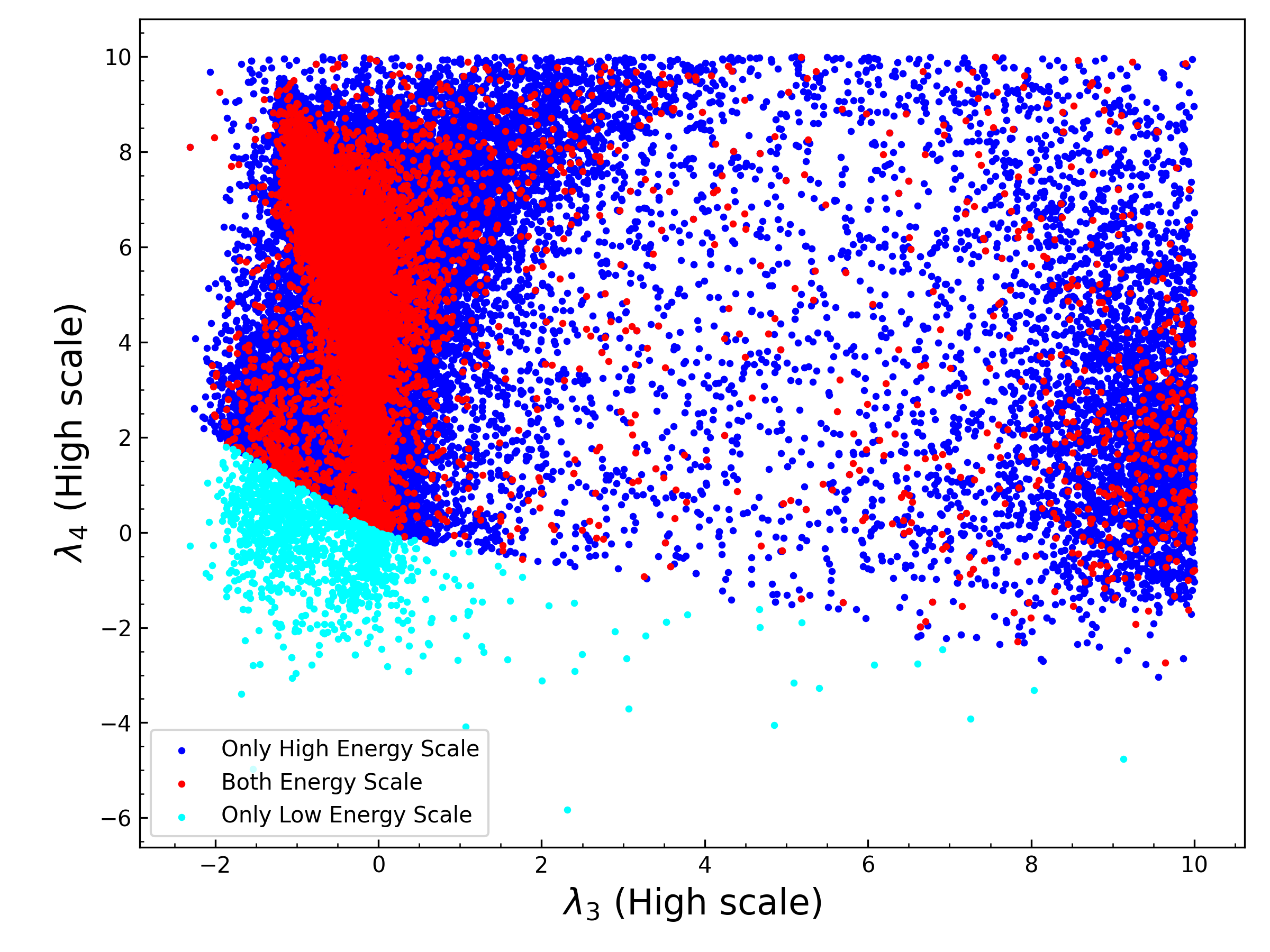}\vspace{-.5cm}
\end{center}
\caption{Positive definiteness constraints for relevant quartic couplings. All points can fulfill the perturbative unitarity bounds. The blue points denote the parameter sets excluded by tree-level BFB constraints at the EW scale but validated by positive definiteness of the RG-improved scalar potential above 10 TeV.  The cyan points denote the parameter sets satisfying tree-level BFB constraints at the EW scale but violating positive definiteness constraints at high scales (large field values). The red dots denote the parameter sets being consistent with positive definiteness constraints at all scales below the Planck scale or the Landau pole scale. For computational simplicity in the numerical scan, the Landau pole scale is defined as the scale where any relevant coupling exceeds 10 during RG evolution.}
\label{fig1}
\end{figure}

We perform a comprehensive numerical scan for those points that satisfy perturbative unitarity and summarize the viable parameter sets in Fig.~\ref{fig1}. For simplicity, any coupling that exceeds 10 during the RG evolution is deemed to have reached its Landau pole. Fig.~\ref{fig1} provides a visual synthesis of these constraints, where:
\begin{itemize}
    \item  Blue points: Parameter sets excluded by tree-level BFB constraints at EW scale but \textit{validated} by RG-improved positive definiteness above $10~\mathrm{TeV}$. This reveals that quantum corrections can \textit{stabilize} otherwise forbidden regions.

    \item  Cyan points: Parameter sets satisfying tree-level BFB constraints at EW scale but \textit{violating} large-field-value positive definiteness constraints. These highlight the danger of relying solely on tree-level analyses.

    \item Red points: Parameter sets consistent with both constraints at all scales, representing the most robust parameter regions.
\end{itemize}

Several critical insights emerge from this analysis:
\begin{itemize}
    \item  The substantial blue region in the $\lambda_1-\lambda_2$ plane (upper panel) demonstrates that a large portion of parameter space previously excluded is \textit{rehabilitated}. This suggests that earlier phenomenological studies that based on tree-level BFB constraints Eq.~(\ref{eq:bfbcond2}) may have been overly restrictive. 

    \item The lower left and right panels show the positive definiteness constraints in the $\lambda_3-\lambda_4$ plane at the EW scale and high energy scale, respectively. The boundary between the red (including cyan) and blue areas in the lower left panel origin from the second formula in Eq.(\ref{eq:bfbcond2}).

        Cyan points concentrated near $\lambda_3 \approx -0.5$ (lower right panel) illustrate how RG-induced sign changes can \textit{destabilize} about one-fifth of tree-level-allowed parameters.

\end{itemize}

\subsection{Constraints in Realistic GM Model}
\label{subsec:fullConstraints}

 Recent measurements and negative searches at the LHC, such as those of the total decay width, Higgs
 strength modifiers and the cross section upper bounds from negative searches of new scalar resonance, impose stringent constraints on the GM model parameter space. 
 We now incorporate all relevant experimental constraints for realistic GM model phenomenology:
\begin{itemize}
\item SM-like Higgs mass and couplings to gauge bosons/fermions ~\cite{ATLAS:kappaWZ,CMS:kappaWZ,ParticleDataGroup:2022pth}.
\item Rare $B$-meson decays sensitive to new physics~\cite{B-physics}.
\item Collider limits on additional Higgs bosons via \texttt{HiggsBounds-5.10.0}~\cite{HiggsBounds1,HiggsBounds2,HiggsBounds3,HiggsBounds4}.
\item SM-like Higgs behavior enforced by \texttt{HiggsSignals-2.6.2}~\cite{HiggsSignals1,HiggsSignals2,HiggsSignals3,HiggsTools}.
\end{itemize}

 Particle spectra and decay properties are computed using specialized tools: \texttt{GMCALC}~\cite{GMCal} for GM-specific calculations and \texttt{SPheno}~\cite{SPheno1,SPheno2} with \texttt{SARAH}~\cite{SARAH1,SARAH2,SARAH3,SARAH4,SARAH5} for general spectrum generation.
Notably, in addition to the free input parameters $\lambda_{1,2,3,4,5}$ and $\sin{\theta}_H$, cubic couplings $M_{1,2}$ now enter our parameter space, which are taken to lie within $[50, 1000]~\mathrm{GeV}$  for realistic models. The mass parameters for doublet and triplets are determined by self-consistency through tadpole conditions.

\begin{table}[htbp]  
\centering  

\begin{tabular}{cccccc}  
\multicolumn{6}{c}{EW scale Inputs (masses in units ${\rm GeV}$)}   \\ 
\hline 
$\lambda_1$ & 0.0430      & $\lambda$ & 0.1722      & $\sigma_4$ & 0.7397  \\  
$\lambda_2$ & $-0.5783$     & $\rho_1$ & $-0.1764$       & $\mu_1$ & 199.7229  \\  
$\lambda_3$ & 0.3561      & $\rho_2$ & $-1.4246$       & $\mu_2$ & 141.2254   \\  
$\lambda_4$ & $-0.1466$     & $\rho_3$ & $-0.4443$       & $\mu_3$ & $-3868.9796$  \\  
$\lambda_5$ & $-0.5231$     & $\rho_4$ & $-1.1566$       & $m_{\phi}$ & 74.5236  \\  
$M_1$ & 282.0764            & $\rho_5$ & 1.4246        & $m_\xi$ & 777.9569   \\  
$M_2$ & 456.1481            & $\sigma_1$ & $-1.1094$  & $m_\chi$ & 789.2162   \\  
$\sin(\theta_H)$ & 0.0859 & $\sigma_2$ & 1.0462    & $\upsilon_{\Delta}$ & 7.4695   \\  
                   &                     & $\sigma_3$ & $-0.2932$  & $\upsilon_{\phi}$ & 246.5897  \\  
\hline   
\end{tabular}  

\vspace{0.2cm}
  
\begin{tabular}{cccc}  
\multicolumn{4}{c}{High energy scale ($2\times10^{18}$GeV)}  \\ 
\hline  
$\lambda$ & 32.9029 & $\sigma_1$ & 1.2304 \\  
$\rho_1$ & 1.5356     & $\sigma_2$ & 0.8382 \\  
$\rho_2$ & $-0.6494$    & $\sigma_3$ & 0.9522 \\  
$\rho_3$ & 1.4619     & $\sigma_4$ & 0.4982 \\  
$\rho_4$ & 1.4289     & $min(\delta_i)$ & 0.0556  \\  
$\rho_5$ & 3.5966     & $max(\delta_i)$ & 2.9472  \\  
\hline  
\end{tabular}  

\vspace{0.2cm}

\begin{tabular}{cccccc}  
\multicolumn{6}{c}{Observeables (masses in units ${\rm GeV}$)}   \\ 
\hline 
$m_{h}$ & 124.8515              & $m_{H^\pm_3}$ & 886.0816          & $\kappa_{WW}$ & 1.0056  \\  
$m_{H_5^0}$ & 887.8551          & $m_{H^{\pm\pm}_5}$ & 884.5776 & $\kappa_{ZZ}$ & 1.0056  \\  
$m_{H_1}$ & 894.3810          & $\kappa_{tt}$ & 0.9980                  & $\kappa_{\gamma\gamma}$ & 1.0553 \\  
$m_{H_3^0}$ & 894.1385      & $\kappa_{bb}$ & 0.9980                & S & $-0.0046$ \\  
$m_{H^\pm_5}$ & 884.6342  & $\kappa_{\tau\tau}$ & 0.9980        & T & $-0.0019$ \\  
\hline   
\end{tabular}  

\vspace{0.2cm}
  
\begin{tabular}{cccc}  
\multicolumn{4}{c}{Observeables}  \\ 
\hline  
U & $-0.0005$                                                                                  & HiggsSignals & $\checkmark$ \\  
$Br(B_s\rightarrow X_{S}\gamma)$ & $3.1470\times{10}^{-4}$  & Unitarity & $\checkmark$ \\  
$Br(B_s\rightarrow \mu^{+}\mu^{-})$ & $3.0978\times{10}^{-9}$ & VS at EW & $\times$ \\  
$Br(B^{+}\rightarrow \tau^{+}\nu)$ & $1.2283\times{10}^{-4}$     & VS at High scale & $\checkmark$ \\  
HiggsBounds & $\checkmark$                                                     & Positive Definiteness & $\checkmark$ \\ 
\hline  
\end{tabular}  

\caption{A Benchmark point for realistic GM model with intermediate magnitude of triplet VEVs. The first sub-table shows the input parameters in (\ref{eq:GMpot}) and the corresponding values in the scalar potential (\ref{eq:Vgeneral}) at the EW scale. The values in the second sub-table are the RGE evolved ones for the corresponding parameters in the first sub-table, with the RGE evolution terminated at $2\times10^{18}$ GeV. We also present some of the low energy experimental observables in the third and fourth sub-tables.  All dimensional parameters listed in the table are given in units of GeV.}
\label{tab:Benchmark1}
\end{table}  

\begin{table}[htbp]
\centering

\begin{tabular}{cccccc}
\multicolumn{6}{c}{EW scale Inputs (masses in units ${\rm GeV}$)}   \\
\hline
$\lambda_1$ & 0.0324    & $\lambda$ & 0.1296    & $\sigma_4$ & 1.0136  \\
$\lambda_2$ & $-0.2289$   & $\rho_1$ & $-4.0926$    & $\mu_1$ & 185.2206  \\
$\lambda_3$ & $-0.5295$   & $\rho_2$ & 2.1180     & $\mu_2$ & 130.9708   \\
$\lambda_4$ & $-0.1788$   & $\rho_3$ & $-1.5168$    & $\mu_3$ & $-756.8336$  \\
$\lambda_5$ & 0.7167    & $\rho_4$ & $-0.9156$    & $m_{\phi}$ & 88.6390  \\
$M_1$ & 261.9415        & $\rho_5$ & $-2.1180$    & $m_\xi$ & 16629.2216   \\
$M_2$ & $-89.1937$        & $\sigma_1$ & $-1.4319$  & $m_\chi$ & 11758.6353   \\
$\sin(\theta_H)$ & $1.6496\times10^{-4}$ & $\sigma_2$ & 1.4334 & $\upsilon_{\Delta}$ & 0.0144   \\
   &   & $\sigma_3$ & -0.3576 & $\upsilon_{\phi}$ & 246.2206  \\
\hline
\end{tabular}

\vspace{0.2cm}

\begin{tabular}{cccc}
\multicolumn{4}{c}{High energy scale ($1.2\times10^{12}$GeV))}  \\
\hline
$\lambda$ & 35.4390     & $\sigma_1$ & 9.0185 \\
$\rho_1$ & 0.5697       & $\sigma_2$ & 8.7407 \\
$\rho_2$ & 0.7392       & $\sigma_3$ & 6.5703 \\
$\rho_3$ & 0.4440       & $\sigma_4$ & 6.0402 \\
$\rho_4$ & 1.5605       & $min(\delta_i)$ & $-0.1404$  \\
$\rho_5$ & $-0.6488$      & $max(\delta_i)$ & 0.5510  \\
\hline
\end{tabular}

\vspace{0.2cm}

\begin{tabular}{cccccc}
\multicolumn{6}{c}{Observeables (masses in units of  ${\rm GeV}$)}   \\
\hline
$m_{h}$ & 125.3545        & $m_{H^\pm_3}$      & 16629.2227    & $\kappa_{WW}$ & 1.0000  \\
$m_{H_5^0}$ & 16627.2640      & $m_{H_5^{\pm\pm}}$   & 16629.2227    & $\kappa_{ZZ}$ & 1.0000  \\
$m_{H_1}$ & 16629.2227      & $\kappa_{tt}$      & 0.9999        & $\kappa_{\gamma\gamma}$ & 1.0424 \\
$m_{H_3^0}$ & 16627.9170      & $\kappa_{bb}$      & 0.9999        & S & $-1.6094\times10^{-5}$ \\
$m_{H^\pm_5}$ & 16627.2640  & $\kappa_{\tau\tau}$ & 0.9999       & T & $-1.2127\times10^{-8}$ \\
\hline
\end{tabular}

\vspace{0.2cm}

\begin{tabular}{cccc}
\multicolumn{4}{c}{Observeables}  \\
\hline
U & $-1.9914\times10^{-7}$                                      & HiggsSignals & $\checkmark$ \\
$Br(B_s\rightarrow X_{S}\gamma)$ & $3.1500\times{10}^{-4}$      & Unitarity & $\checkmark$ \\
$Br(B_s\rightarrow \mu^{+}\mu^{-})$ & 3.0376$\times{10}^{-9}$   & VS at EW & $\times$ \\
$Br(B^{+}\rightarrow \tau^{+}\nu)$ & 1.1257$\times{10}^{-4}$    & VS at High scale & $\checkmark$ \\
HiggsBounds & $\checkmark$                                      & Positive Definiteness & $\checkmark$ \\
\hline
\end{tabular}

\caption{Another benchmark point for realistic GM model with tiny VEVs for triplets. The settings of this table are exactly the same as those of Table \ref{tab:Benchmark1}.}
\label{tab:Benchmark2}
\end{table}

Once a set of parameters that satisfy perturbative unitarity bound and experimental constraints are specified, we can evolve the input parameters via renormalization group equations from
the EW scale to the Planck scale or to the Landau pole scale and derive the positive definiteness constraints. Two benchmark points are present to illuminate the positive definiteness constraints:

\textbf{Benchmark 1 (Intermediate $v_\Delta$):} Table~\ref{tab:Benchmark1} shows a representative point with $v_\Delta \approx 7.5~\text{GeV}$. Key features include:
\begin{itemize}
\item \textit{EW scale parameters:} Negative $\lambda_2$ and $\lambda_4$ would traditionally violate BFB constraints.
\item \textit{High-scale (large field value) evolution:} The quartic coupling $\lambda$ grows to 32.9 while $\rho_5$ reaches 3.6, with significant custodial breaking ($\max\delta_i \approx 2.95$).
\item \textit{Observables:} It predicts 125 GeV Higgs with SM-like couplings ($\kappa_{VV} \approx 1.006$), heavy additional scalars ($m_{H^\pm}, m_{H_3^0},\cdots \sim 884{\rm GeV}-894$ {\rm GeV}).
\item \textit{Vacuum Stability:} Although tree-level BFB constraints are violated, ordinary custodial symmetry preserving EW vacuum is stable because of the positive definiteness of the scalar potential at large field value regions. The orders-of-magnitude coupling growth at large field value $2\times10^{18}~\mathrm{GeV}$ and custodial breaking ($\max|\delta_i| = 2.95$) exemplify that quantum effects can reshape the potential landscape.
\end{itemize}
This demonstrates how RG evolution can stabilize initially problematic parameter sets.

\textbf{Benchmark 2 (Tiny $v_\Delta$):} Table~\ref{tab:Benchmark2} presents a case with negligible triplet VEVs ($v_\Delta < 1~\mathrm{GeV}$). Key features include:
\begin{itemize}
\item \textit{Spectrum:} Heavy scalars $\sim 16.6$ TeV with near-perfect SM alignment ($\kappa_i \approx 1$).
\item \textit{Precision tests:} Negligible oblique corrections ($S, T, U \sim 10^{-5}$).
\item \textit{Vacuum Stability:} Tree-level BFB constraints are violated but ordinary custodial symmetry preserving EW vacuum is stable because of the positive definiteness of the scalar potential at large field value regions.
\end{itemize}
Both benchmarks satisfy all experimental constraints while illustrating the core finding: tree-level BFB constraints fails to capture the positive definiteness requirements of the effective potential.

\begin{figure}[htbp]
\begin{center}
\includegraphics[width=7.5 cm]{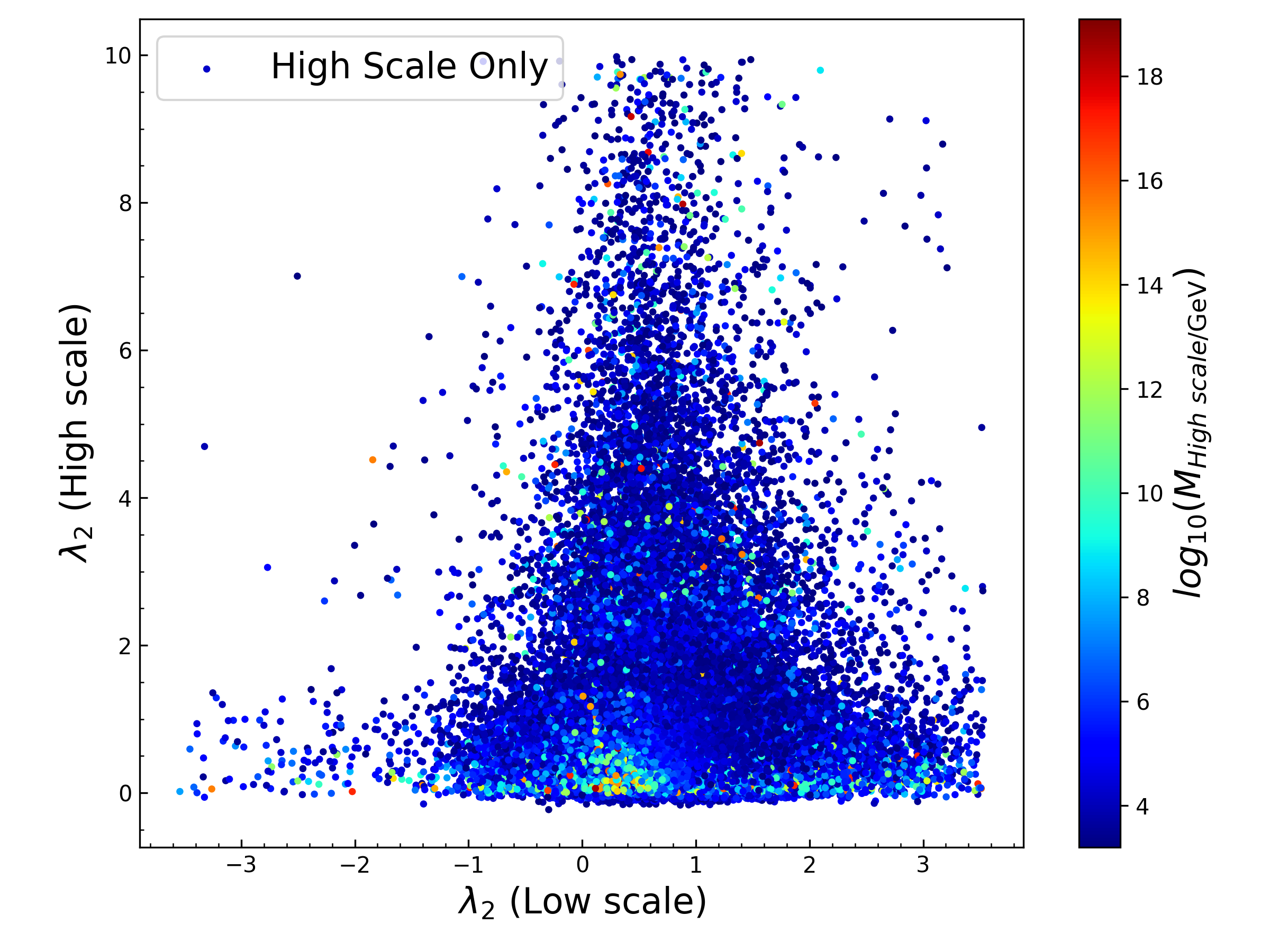}
    \includegraphics[width=7.5 cm]{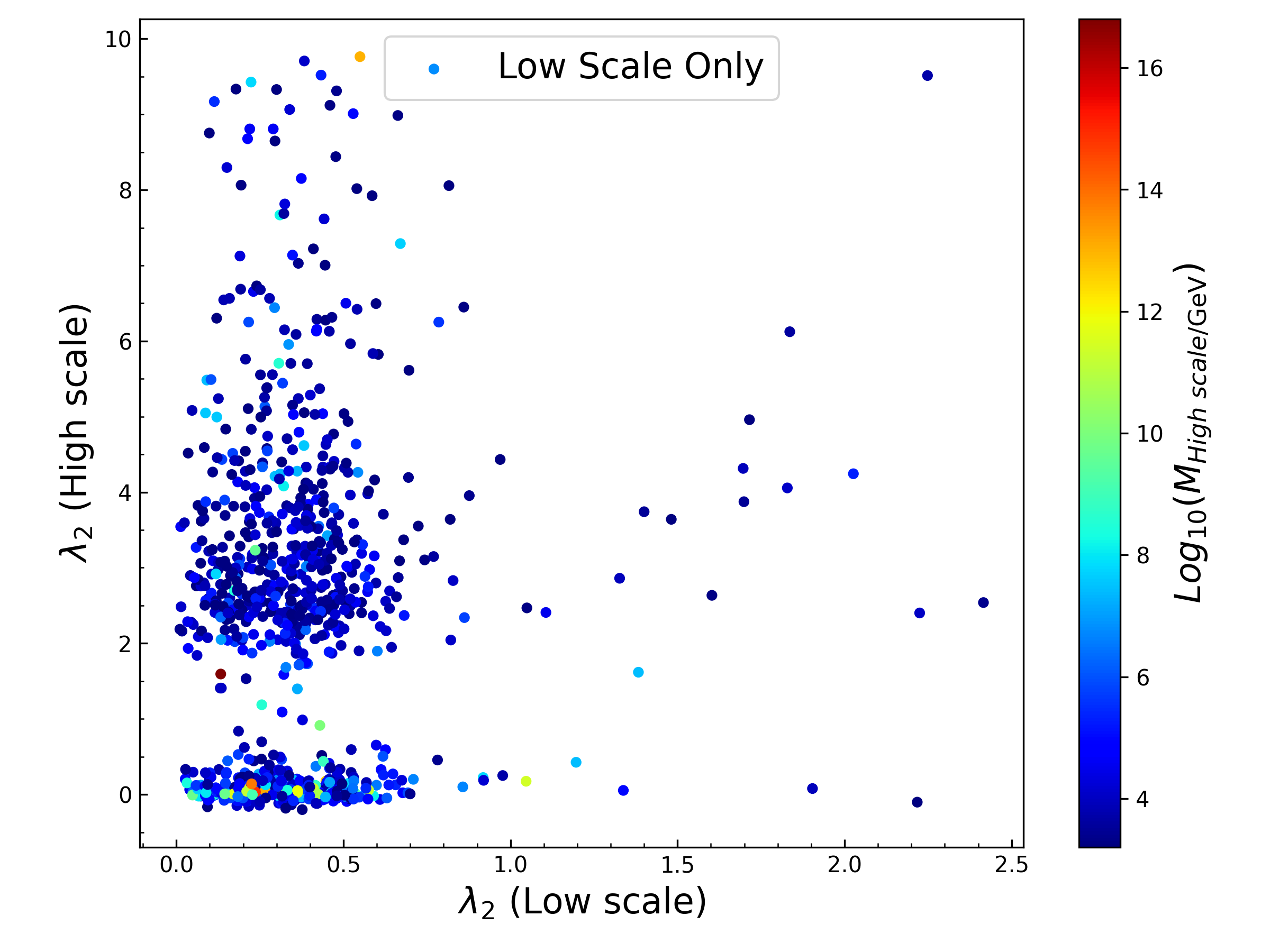}
  \\
    \includegraphics[width=7.5 cm]{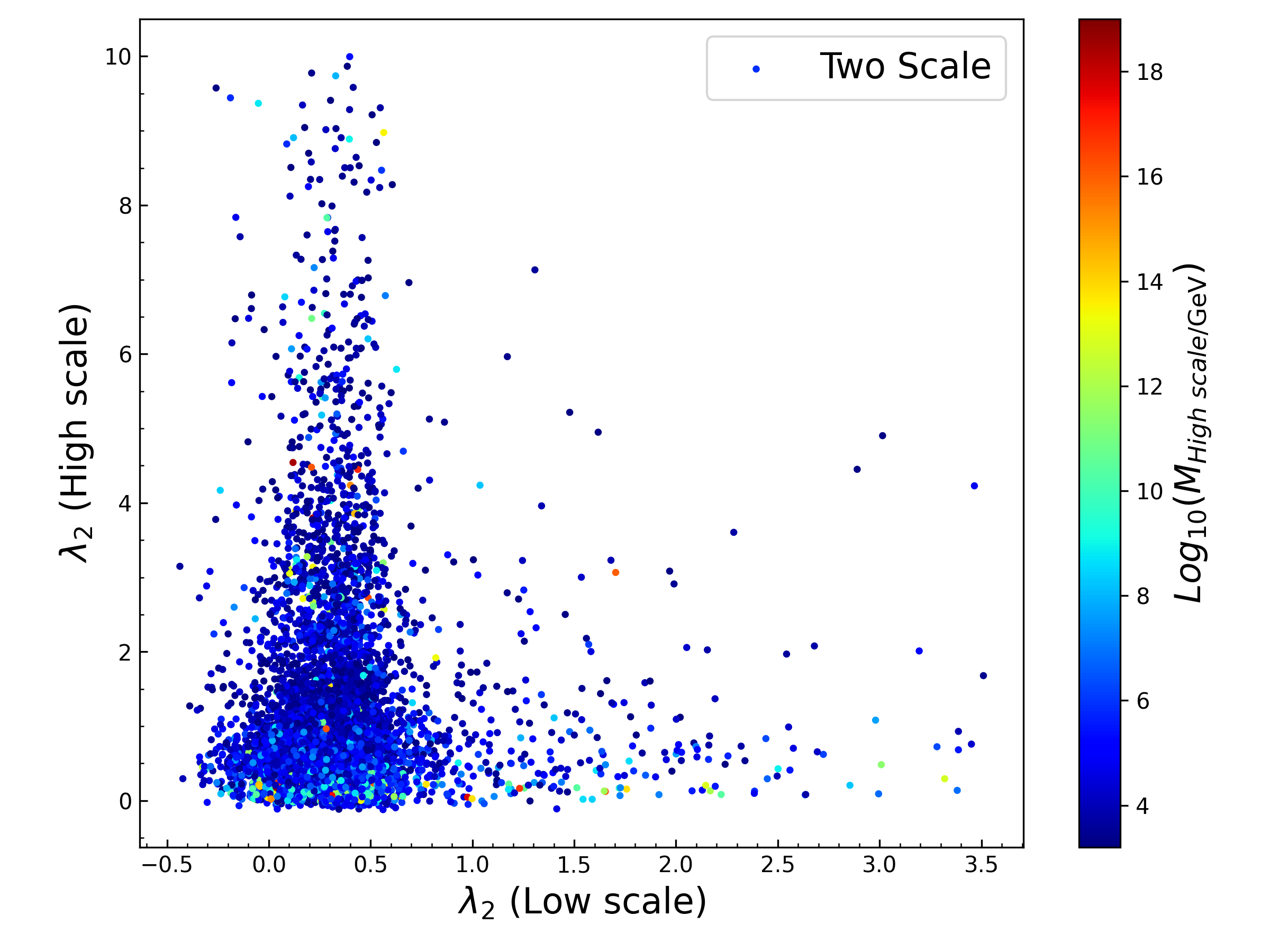}
    \includegraphics[width=7.5 cm]{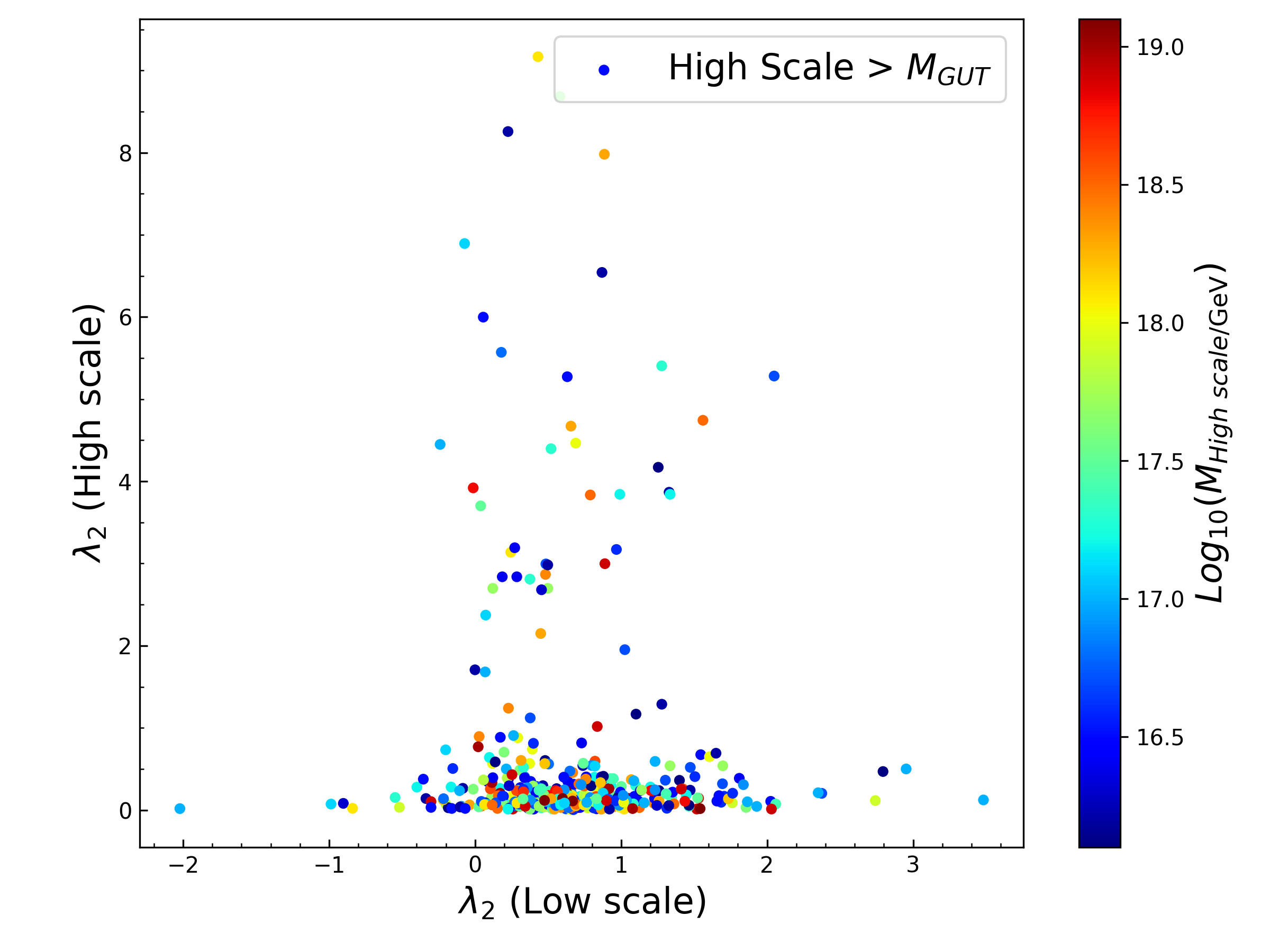}
\end{center}
\caption{ Evolution of quartic coupling $\lambda_2$ values: EW scale vs. high scale after renormalization group evolution. Colors indicate the maximum scale achieved via RG evolution before encountering a Landau pole or reaching the Planck scale.
In the panels (a-c), we show the parameter sets satisfying positive definiteness above 10 TeV, only at the EW scale, and at all scales below the Planck/Landau pole scale, respectively. Panel (d) shows the parameter sets satisfying positive definiteness up to the Planck scale (without encountering a Landau pole below it)}
\label{fig2}
\end{figure}
In Fig.~\ref{fig2}, we present a comprehensive scan of parameter space passing all experimental/theoretical constraints except tree-level BFB constraints. Different color correspond to different maximum scale (reaching the corresponding Landau pole scale for some quartic coupling or reaching the Planck scale) that each input parameter point can achieve by renormalization group evolution. We choose the $\lambda_2$ quartic coupling to illustrate the effects of renormalization group evolution on positive definiteness constraints. The $\lambda_2$ values at EW scale versus high scale reveal:
\begin{itemize}
\item \textbf{Panel (a):} Region satisfying positive definiteness above $\phi \gtrsim 10~\text{TeV}$ is substantially larger than the tree-level BFB region.
\item \textbf{Panel (b):} Parameters stable only at EW scale form a narrow boundary zone.
\item \textbf{Panel (c):} The subset maintaining positive definiteness at all scales is highly constrained.
\item \textbf{Panel (d):} A small portion of points can maintain positive definiteness up to the $M_{\text{Planck}}$ scale without Landau poles, though quantum gravity effects may modify this conclusion near $M_{\mathrm{Pl}}$.
\end{itemize}

Critical implications:
\begin{enumerate}
\item The significant area difference between Panels (a) and (c) represents theoretically viable parameter space previously excluded by tree-level BFB constraints. In fact, almost one-third of the parameter space previously excluded is reinstated--these regions should be prioritized in future phenomenological studies of the GM model. In addition, almost one-fifth of tree-level-allowed points become unstable at high scales, which reveal instabilities induced by  RG evolution for apparently stable EW-scale parameters.

\item  The quantification of custodial violation through $\delta_i$ parameters reveals that when $|\delta_i|$ is non-negligible (typically, when $|\delta_i|\gtrsim 0.1$), tree-level stability conditions become unreliable in a large majority of sample cases. This establishes $|\delta_i|$ as a key diagnostic for the reliability of tree-level BFB constraints.

\item These reinstated parameters predict testable phenomena: modified Higgs couplings, heavy scalar production, and exotic decays.

\end{enumerate}
This demonstrates that the traditional BFB criterion \textit{underestimates} viable parameter space while simultaneously \textit{overlooking} potential instabilities - both resolved by our RG-improved positive definiteness approach.

\begin{figure}[htbp]  
\begin{center}  
\includegraphics[width=7.5 cm]{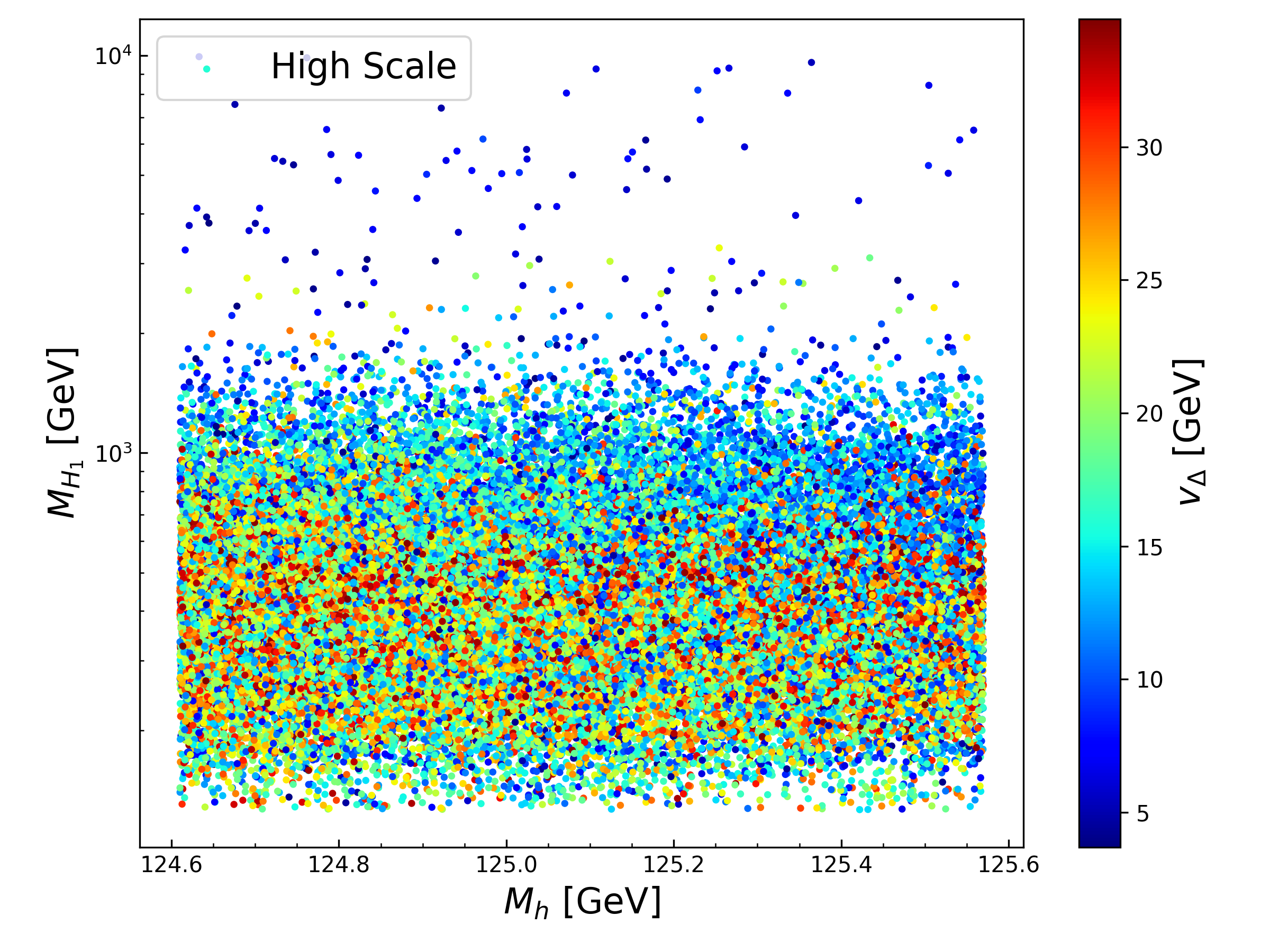}
    \includegraphics[width=7.5 cm]{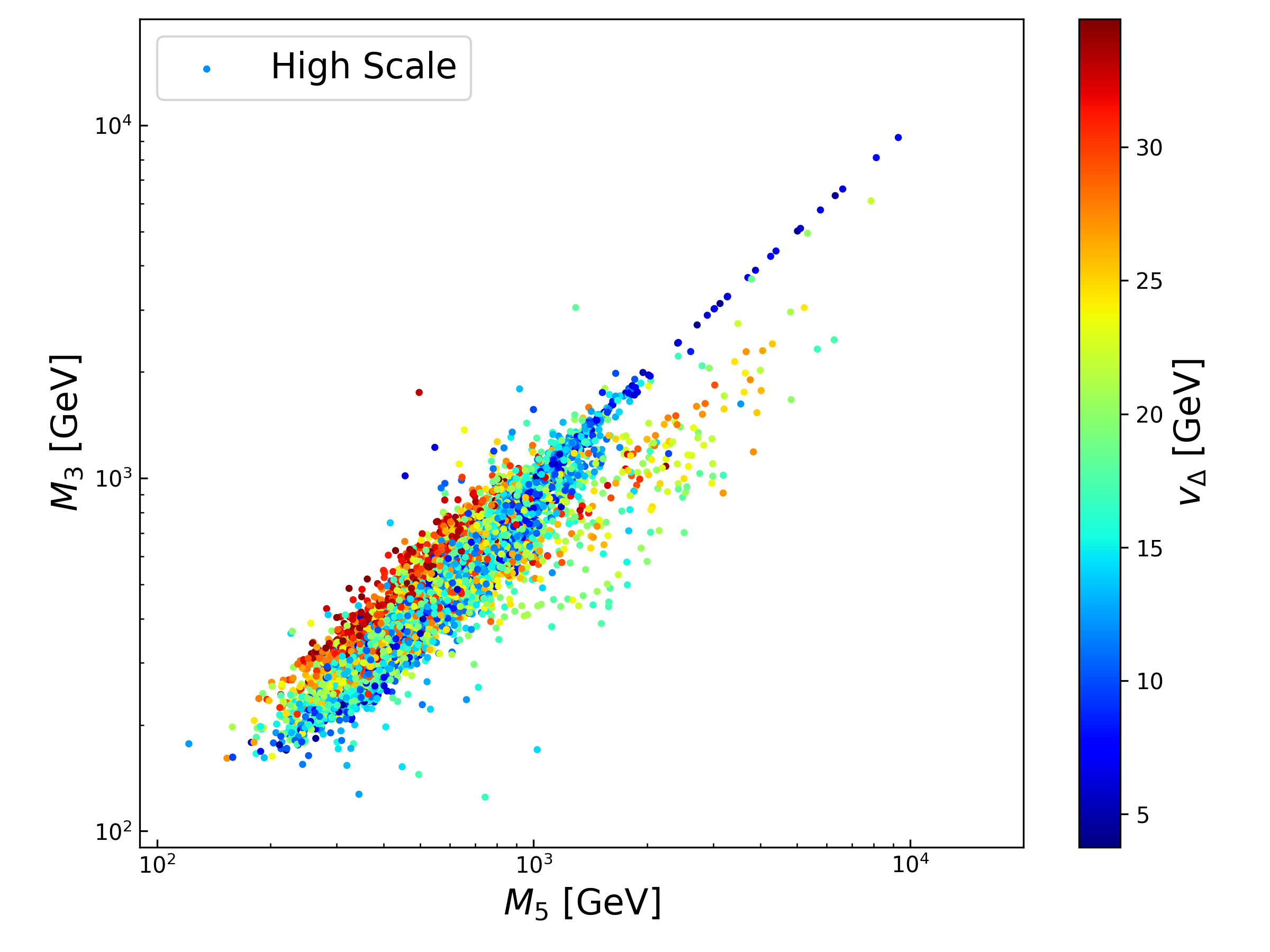}
  \\
    \includegraphics[width=7.5 cm]{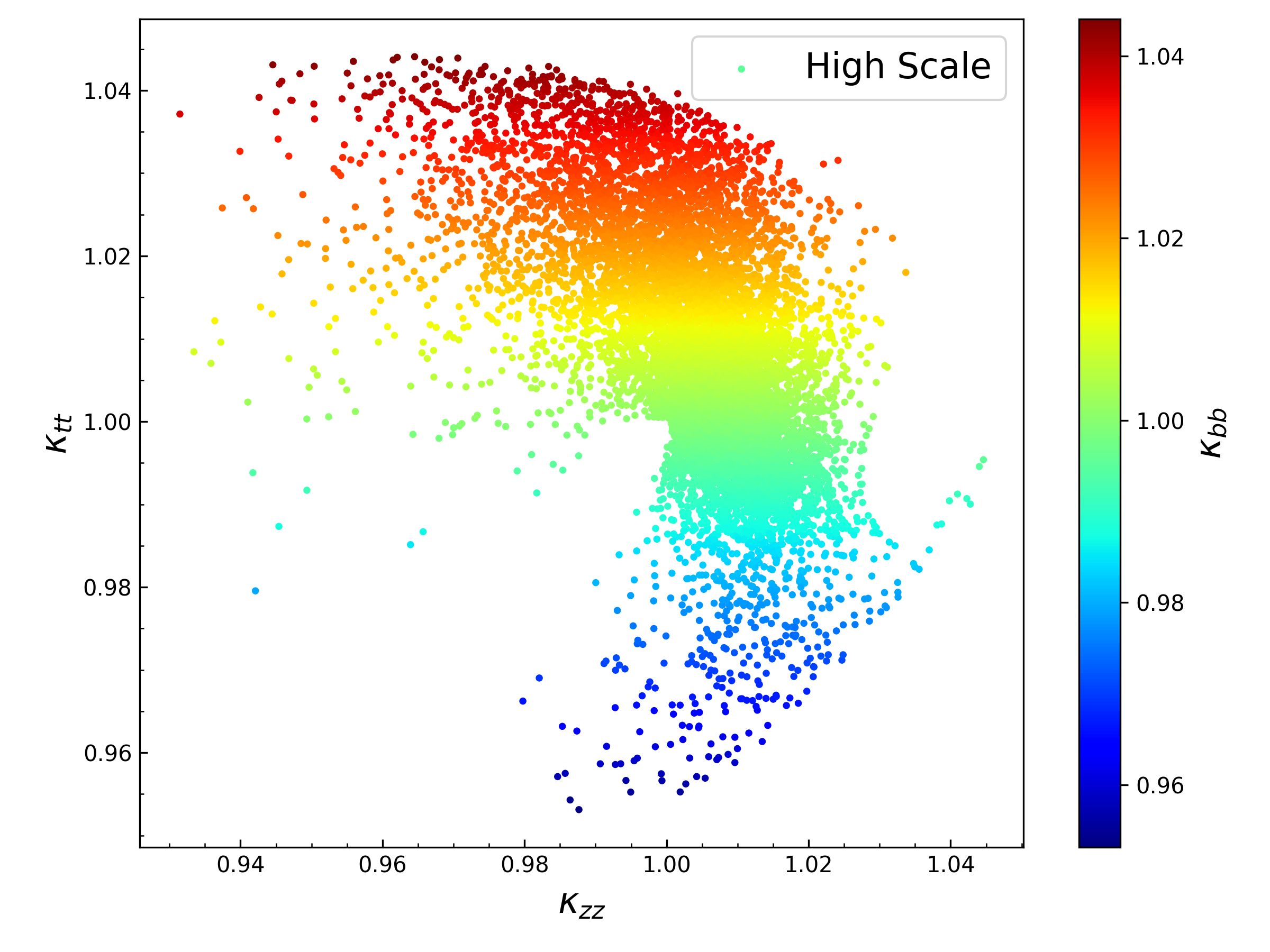}
    \includegraphics[width=7.5 cm]{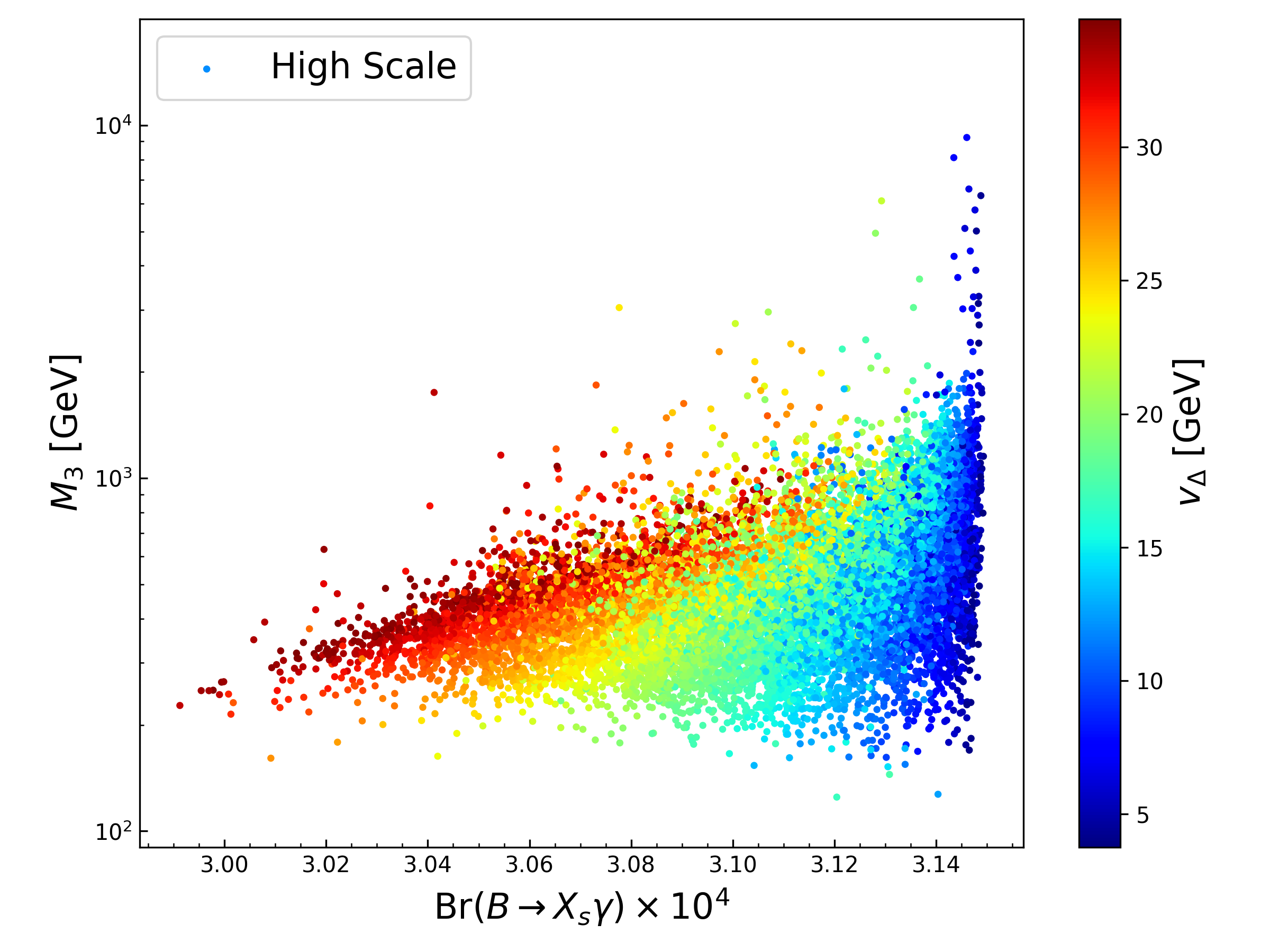}
\end{center}
\caption{ The values of typical physical parameter observables predicted by the GM model, after imposing the new positive definiteness constraints and all other experimental/theoretical constraints (except the tree-level BFB constraints).
Panels (a) and (b) show the correlations of triplets VEVs $v_\Delta$ with the masses of the custodial singlets $h,~H_1$, the custodial triplet $H_3$, and the quintuplet $H_5$, respectively. The effective couplings of the SM-like 125 GeV Higgs boson and third generation quarks and the Z vector boson are presented in the panel (c).
The panel (d) show the branching ratios of the rare B decays $BR(B\ra X_s \gamma)$ versus $V_\Delta$ and $M_3$.} 
\label{fig3}  
\end{figure}

 The surviving values of the quartic couplings $\lambda_i$ can be translated into physical observables of the realistic GM model, such as the masses of the custodial singlets $h$ and $H_1$, the custodial triplet $H_3$, and the quintuplet $H_5$, as given by Eq.~(\ref{multiplet:masses}). The relevant correlations are illustrated in the upper panels of Fig.~\ref{fig3}, where the colors correspond to the triplet VEVs $v_{\Delta}$. The upper panels reveal that large masses of the new scalars $H_{1,3,5}$ favor small triplet VEVs, while smaller masses are associated with larger triplet VEVs. In fact, over a significant portion of the sample points, the masses of the new scalars $H_1$ and $H_3$ remain below 3 TeV, particularly when the triplet VEVs exceed 20 GeV. Similarly, the masses of $H_5$ are typically below 3 TeV in a large fraction of the sample points, especially when $v_{\Delta} > 30$ GeV.

The 125 GeV SM-like Higgs boson predicted by the GM model must be consistent with current ATLAS and CMS measurements. In the lower left panel of Fig.~\ref{fig3}, we present the effective coupling modifiers of the SM-like Higgs relative to the SM, specifically the modifiers for couplings of SM-like Higgs $h$ to third-generation quarks ($\kappa_{tt}$, $\kappa_{bb}$) and to the $Z^0$ gauge boson ($\kappa_{ZZ}$). The panel clearly shows that the couplings of the 125 GeV Higgs boson to fermions and gauge bosons closely approach their SM values. The parameter points shown in Fig.~\ref{fig3}, which survive all current experimental constraints---including those on the effective couplings of the SM-like Higgs---can be further tested in future collider experiments such as the HL-LHC~\cite{HL-LHC1,HL-LHC2} and the CEPC~\cite{CEPC1,CEPC2,CEPC3}.

The couplings of scalars to fermions in the GM model have exactly the same structure as those in the Type-I two-Higgs-doublet model with appropriate replacements~\cite{GM56}. The dominant new-physics contributions to $B$-physics observables therefore arise solely from diagrams involving scalar couplings to the top quark. Since the charged triplet $H_3$ partially originates from the SM doublet, it can couple to both up- and down-type quarks, leading to flavor-violating processes such as $b \to s$ transitions via diagrams involving the $H_3^+ \bar{t}b$ coupling. Note that the custodial quintuplet states do not couple to fermions, as they contain no $SU(2)_L$ doublet components. Therefore, $b \to s$ transition processes depend only on the charged triplet mass $M_3$ and the mixing angle $\beta$ (between $\phi^\pm$ and $(\chi^\pm + \xi^\pm)$ in Eq.(\ref{H3:combination})). Experimental data on $B$-physics can be used to constrain these parameters, imposing an upper bound on $v_{\Delta}$ as a function of $M_3$~\cite{GM56}. In the lower right panel of Fig.~\ref{fig3}, the predicted branching ratios for rare $B$-meson decays are shown (using $B \to X_s \gamma$ as an example). It is evident from the panel that very small triplet scalar masses $M_3 \lesssim \mathcal{O}(100)\ \mathrm{GeV}$ or large $v_{\Delta} \gtrsim \mathcal{O}(50)\ \mathrm{GeV}$ are ruled out by the measured branching ratios of rare $B$ decays.


\section{Conclusions}
\label{sec:conclusions}

The tree-level BFB constraints for the GM model can set stringent constraints on the input parameters in phenomenological studies. However, the landscape of the effective scalar potential can be quite different from the tree-level scalar potential, necessitating the reinvestigation of the theoretical BFB/positivity constraints for the scalar potential.

Using a one-loop RG-improved effective potential and new criteria for positive definiteness of homogeneous polynomials with multiple variables (necessary due to custodial symmetry breaking
effects from loops), we numerically analyze the positive definiteness constraints of the effective potential in the GM model to ensure the absence of deeper vacua in regions with large field values. Our systematic investigation establishes that vacuum stability in the GM model is fundamentally reshaped by renormalization group evolution--a conclusion with profound implications for both theoretical consistency and phenomenological viability. Traditional tree-level BFB conditions provide an incomplete description of stability, particularly in large field value regions where quantum corrections reshape the landscape of the scalar potential.

We should note that several critical questions demand further studies:
\begin{itemize}
    \item The possibility of \textit{metastability} vacuum: Regions with unstable EW vacua but lifetimes exceeding the universe's age ($\tau > 10^{10}$ yr) remain phenomenologically viable. It was noted in~\cite{GM60} that GM model has a rich vacuum structure and many false vacua can coexist with the true vacuum, leading to multiple-step transitions to the true minimum. So, unless we fully understand the vacuum structure of the effective potential at large field value regions, naive estimation of the lifetime for meta-stable EW vacuum by calculating its transition rate to the true vacuum is always unreliable.  Calculating tunneling rates in these multi-minima potentials presents both computational and conceptual challenges. We would like to leave such studies for future works.

    \item \textit{Two-loop RG effects} may alter stability boundaries: Our one-loop analysis provides a robust foundation, but precision studies should incorporate higher-order corrections.

    \item The \textit{experimental accessibility} of reinstated regions warrants urgent investigation: Some enhanced exotic Higgs (such as the doubly charged $H_5^{\pm\pm}$ scalar) production rates and distinctive Higgs coupling patterns predicted in these regions can possibly be probed at the HL-LHC or CEPC.
\end{itemize}

\noindent In summary, this work demonstrates that quantum corrections dramatically reconfigure the Georgi-Machacek vacuum landscape. Taking into account quantum effects, we reveal new theoretical parameter regions for exploration. Numerical analysis, based on our new criteria for positive definiteness of homogeneous polynomials with multiple variables, not only revises the GM model's viability map but also provides a methodological template for stability studies in other extended Higgs models.


\begin{acknowledgments}
This work was supported by the National Natural Science Foundation of China (NNSFC) under grant Nos.12075213,12335005 and 12447167, by the Natural Science Foundation for Distinguished Young Scholars of Henan Province under grant number 242300421046, by the the Joint Fund of Henan Province Science and Technology R$\&$D Program No.225200810092 and 225200810030, by the Startup Research Fund of Henan Academy of Sciences No.231820011, by the Basic Research Fund of Henan Academy of Sciences No.240620006.
\end{acknowledgments}

\appendix

\titleformat{\section}{\normalfont\Large\bfseries}{Appendix \thesection}{1em}{}
\section{One-loop $\beta$-functions for GM model}
\label{app:A}
In our numerical studies, we need to evaluate the RGE evolution of various couplings. The one-loop RGE for the gauge couplings $g_{1,2,3}$, the Yukawa couplings $Y_{u,d,e}$ and quartic couplings $\lambda$, $\rho_{1,2,3,4,5}$,$\sigma_{1,2,3,4}$ can be calculated by the general formulas in~\cite{Machacek:1983tz, Machacek:1983fi, Machacek:1984zw,oneloop:RGE} or by the package SARAH~\cite{SARAH1,SARAH2}, which is a Mathematica package that can generate two-loop RGEs for supersymmetric and non-supersymmetric models. Analytical expressions of the RGE for GM model had already been presented in various papers~\cite{GM44,GM45,Kundu:2021pcg,GM64}.  We recalculate the RGEs with SARAH to check the results in the literatures and present here the complete one-loop $\beta$-functions for later convenience:
\beqa
\beta_{g_1} & =&
~\frac{47}{10} g_{1}^{3},~~~
\beta_{g_2} =
-\frac{13}{6} g_{2}^{3},~~~
\beta_{g_3}  =
-7 g_{3}^{3},~~~~~~~~~~~~~~~~~~~~~~~~~~~~~~~~~~~~~~~~~
\eeqa
\beqa
\beta_{Y_u} & = &
-\frac{3}{2} \Big(- {Y_u  Y_{u}^{\dagger}  Y_u}  + {Y_u  Y_{d}^{\dagger}  Y_d}\Big)\nonumber \\
&+&Y_u \Big(3 \mbox{Tr}\Big({Y_d  Y_{d}^{\dagger}}\Big)  + 3 \mbox{Tr}\Big({Y_u  Y_{u}^{\dagger}}\Big)  -8 g_{3}^{2}  -\frac{17}{20} g_{1}^{2}  -\frac{9}{4} g_{2}^{2}  + \mbox{Tr}\Big({Y_e  Y_{e}^{\dagger}}\Big)\Big),\\ 
\beta_{Y_d} & =&
\frac{1}{4} \Big(6 \Big(- {Y_d  Y_{u}^{\dagger}  Y_u}  + {Y_d  Y_{d}^{\dagger}  Y_d}\Big)\nonumber \\
&-& Y_d \Big(-12 \mbox{Tr}\Big({Y_d  Y_{d}^{\dagger}}\Big)  -12 \mbox{Tr}\Big({Y_u  Y_{u}^{\dagger}}\Big)  + 32 g_{3}^{2}  -4 \mbox{Tr}\Big({Y_e  Y_{e}^{\dagger}}\Big)  + 9 g_{2}^{2}  + g_{1}^{2}\Big)\Big),\\ 
\beta_{Y_e} & =&
\frac{3}{2} {Y_e  Y_{e}^{\dagger}  Y_e}  + Y_e \Big(3 \mbox{Tr}\Big({Y_d  Y_{d}^{\dagger}}\Big)  + 3 \mbox{Tr}\Big({Y_u  Y_{u}^{\dagger}}\Big)  -\frac{9}{4} g_{1}^{2}  -\frac{9}{4} g_{2}^{2}  + \mbox{Tr}\Big({Y_e  Y_{e}^{\dagger}}\Big)\Big), 
\eeqa
and
\beqa
\beta_{{\lambda}_{1}} & =&
\frac{27}{200} g_{1}^{4} +\frac{9}{20} g_{1}^{2} g_{2}^{2} +\frac{9}{8} g_{2}^{4} -\frac{9}{5} g_{1}^{2} {\lambda}_{1} -9 g_{2}^{2} {\lambda}_{1} +24 {\lambda}_{1}^{2} +3 {\sigma}_1^{2} +3 {\sigma}_1 {\sigma}_2 +\frac{5}{4} {\sigma}_2^{2} +6 {\sigma}_3^{2} \notag\\
&+&2 |{\sigma}_4|^2 +12 {\lambda}_{1} \mbox{Tr}\Big({Y_d  Y_{d}^{\dagger}}\Big) +4 {\lambda}_{1} \mbox{Tr}\Big({Y_e  Y_{e}^{\dagger}}\Big) +12 {\lambda}_{1} \mbox{Tr}\Big({Y_u  Y_{u}^{\dagger}}\Big) -6 \mbox{Tr}\Big({Y_d  Y_{d}^{\dagger}  Y_d  Y_{d}^{\dagger}}\Big) \notag\\
&-& 2 \mbox{Tr}\Big({Y_e  Y_{e}^{\dagger}  Y_e  Y_{e}^{\dagger}}\Big) -6 \mbox{Tr}\Big({Y_u  Y_{u}^{\dagger}  Y_u  Y_{u}^{\dagger}}\Big),
\eeqa
\beqa
\beta_{{\sigma}_1} & =&
\frac{27}{25} g_{1}^{4} -\frac{18}{5} g_{1}^{2} g_{2}^{2} +6 g_{2}^{4} -\frac{9}{2} g_{1}^{2} {\sigma}_1 -\frac{33}{2} g_{2}^{2} {\sigma}_1 +12 {\lambda}_{1} {\sigma}_1 +16 {\rho}_{1} {\sigma}_1 +12 {\rho}_{2} {\sigma}_1 +4 {\sigma}_1^{2} \notag\\
&+& 4 {\lambda}_{1} {\sigma}_2 +6 {\rho}_{1} {\sigma}_2 +2 {\rho}_{2} {\sigma}_2 +{\sigma}_2^{2}+12 {\rho}_{4} {\sigma}_3 +4 {\rho}_{5} {\sigma}_3 +2 |{\sigma}_4|^2 +6 {\sigma}_1 \mbox{Tr}\Big({Y_d  Y_{d}^{\dagger}}\Big) \notag\\
&+&2 {\sigma}_1 \mbox{Tr}\Big({Y_e  Y_{e}^{\dagger}}\Big) +6 {\sigma}_1 \mbox{Tr}\Big({Y_u  Y_{u}^{\dagger}}\Big), 
\eeqa
\beqa
\beta_{{\sigma}_2} & =&
\frac{36}{5} g_{1}^{2} g_{2}^{2} -\frac{9}{2} g_{1}^{2} {\sigma}_2 -\frac{33}{2} g_{2}^{2} {\sigma}_2 +4 {\lambda}_{1} {\sigma}_2 +4 {\rho}_{1} {\sigma}_2 +8 {\rho}_{2} {\sigma}_2 +8 {\sigma}_1 {\sigma}_2 +4 {\sigma}_2^{2} +4 |{\sigma}_4|^2 \nn\\
&+&6 {\sigma}_2 \mbox{Tr}\Big({Y_d  Y_{d}^{\dagger}}\Big) +2 {\sigma}_2 \mbox{Tr}\Big({Y_e  Y_{e}^{\dagger}}\Big) +6 {\sigma}_2 \mbox{Tr}\Big({Y_u  Y_{u}^{\dagger}}\Big),
\eeqa
\beqa
\beta_{{\sigma}_3} & =&
3 g_{2}^{4} +6 {\rho}_{4} {\sigma}_1 +2 {\rho}_{5} {\sigma}_1 +3 {\rho}_{4} {\sigma}_2 +{\rho}_{5} {\sigma}_2 -\frac{9}{10} g_{1}^{2} {\sigma}_3 -\frac{33}{2} g_{2}^{2} {\sigma}_3 +12 {\lambda}_{1} {\sigma}_3 +20 {\rho}_{3} {\sigma}_3 \notag\\
&+& 8 {\sigma}_3^{2} +4 |{\sigma}_4|^2 +6 {\sigma}_3 \mbox{Tr}\Big({Y_d  Y_{d}^{\dagger}}\Big) +2 {\sigma}_3 \mbox{Tr}\Big({Y_e  Y_{e}^{\dagger}}\Big) +6 {\sigma}_3 \mbox{Tr}\Big({Y_u  Y_{u}^{\dagger}}\Big),\eeqa
\beqa \beta_{{\sigma}_4} & =&
-\frac{27}{10} g_{1}^{2} {\sigma}_4 -\frac{33}{2} g_{2}^{2} {\sigma}_4 +4 {\lambda}_{1} {\sigma}_4 +4 {\rho}_{4} {\sigma}_4 -2 {\rho}_{5} {\sigma}_4 +4 {\sigma}_1 {\sigma}_4 +4 {\sigma}_2 {\sigma}_4 +8 {\sigma}_3 {\sigma}_4 \notag\quad\quad\\
&+& 6 {\sigma}_4 \mbox{Tr}\Big({Y_d  Y_{d}^{\dagger}}\Big) +2 {\sigma}_4 \mbox{Tr}\Big({Y_e  Y_{e}^{\dagger}}\Big) +6 {\sigma}_4 \mbox{Tr}\Big({Y_u  Y_{u}^{\dagger}}\Big),\eeqa 
\beqa
\beta_{{\rho}_{1}} & =&
15 g_{2}^{4}  -24 g_{2}^{2} {\rho}_{1}  + 24 {\rho}_{1} {\rho}_{2}  + 28 {\rho}_{1}^{2}  + 2 {\sigma}_1^{2}  + 2 {\sigma}_1 {\sigma}_2  + 3 {\rho}_{5}^{2}  + 4 {\rho}_{4} {\rho}_{5}  + 6 {\rho}_{2}^{2}  + 6 {\rho}_{4}^{2}  \notag\quad\\
&-&\frac{36}{5} g_{1}^{2} \Big(g_{2}^{2} + {\rho}_{1}\Big) + \frac{54}{25} g_{1}^{4}, \eeqa  
\beqa\beta_{{\rho}_{2}} & = &
18 {\rho}_{2}^{2}  -24 g_{2}^{2} {\rho}_{2}  + 24 {\rho}_{1} {\rho}_{2}  -2 {\rho}_{5}^{2}  -6 g_{2}^{4}  + \frac{36}{5} g_{1}^{2} \Big(2 g_{2}^{2}  - {\rho}_{2} \Big) + {\sigma}_2^{2},~~~~~~~~~~~~~~~~~\eeqa
\beqa\beta_{{\rho}_{3}} & =&
2 \Big(-12 g_{2}^{2} {\rho}_{3}  + 22 {\rho}_{3}^{2}  + 2 {\rho}_{4} {\rho}_{5}  + 2 {\sigma}_3^{2}  + 3 g_{2}^{4}  + 3 {\rho}_{4}^{2}  + {\rho}_{5}^{2}\Big),~~~~~~~~~~~~~~~~~~~~~~~~~\eeqa 
\beqa
\beta_{{\rho}_{4}} & =&
6 g_{2}^{4} -\frac{18}{5} g_{1}^{2} {\rho}_{4} -24 g_{2}^{2} {\rho}_{4} +16 {\rho}_{1} {\rho}_{4} +12 {\rho}_{2} {\rho}_{4} +20 {\rho}_{3} {\rho}_{4} +8 {\rho}_{4}^{2} +4 {\rho}_{1} {\rho}_{5} +4 {\rho}_{2} {\rho}_{5} \notag \\
&+&4 {\rho}_{3} {\rho}_{5} +2 {\rho}_{5}^{2} +4 {\sigma}_1 {\sigma}_3 +2 {\sigma}_2 {\sigma}_3 +2 |{\sigma}_4|^2,\eeqa
\beqa
\beta_{{\rho}_{5}} & = &
-24 g_{2}^{2} {\rho}_{5}  + 2 {\rho}_{5} \Big(2 {\rho}_{1}  + 4 {\rho}_{3}  + 5 {\rho}_{5}  + 8 {\rho}_{4} \Big) -2 |{\sigma}_4|^2  + 6 g_{2}^{4}  -\frac{18}{5} g_{1}^{2} {\rho}_{5},\quad\quad\quad\quad
\eeqa
with
\beqa
\f{d}{d\ln\mu} \tl{\lambda}= \f{1}{16\pi^2}\beta_{\tl{\lambda}}~,
\eeqa
for any coupling $\tl{\lambda}$. It should be noted that the $\beta$ function for $U(1)_Y$ gauge coupling $g_1$ differs by a normalization factor  $3/5$ from that given in~\cite{GM44}, in which the GUT normalization $5g_1^2=3({g^\prime})^2$ is adopted.

For completeness, the beta functions for dimensional couplings are also list here
\beqa
\beta_{\mu} & = &
-\frac{27}{10} g_{1}^{2} \mu -\frac{21}{2} g_{2}^{2} \mu +4 {\lambda}_{1} \mu +4 \mu {\sigma}_1 +6 \mu {\sigma}_2 +\Big(-2 {\mu}_{3}  + 4 {\mu}_{1} \Big){\sigma}_4^* +6 \mu \mbox{Tr}\Big({Y_d  Y_{d}^{\dagger}}\Big) \nonumber \\
&+& 2 \mu \mbox{Tr}\Big({Y_e  Y_{e}^{\dagger}}\Big) +6 \mu \mbox{Tr}\Big({Y_u  Y_{u}^{\dagger}}\Big),\\ 
\beta_{{\mu}_{1}} & =&
-\frac{9}{10} g_{1}^{2} {\mu}_{1} -\frac{21}{2} g_{2}^{2} {\mu}_{1} +4 {\lambda}_{1} {\mu}_{1} -2 {\mu}_{3} {\sigma}_2 +8 {\mu}_{1} {\sigma}_3 +8 \mu {\sigma}_4 +8 \mu^* {\sigma}_4^* +6 {\mu}_{1} \mbox{Tr}\Big({Y_d  Y_{d}^{\dagger}}\Big)\nn \\
&+& 2 {\mu}_{1} \mbox{Tr}\Big({Y_e  Y_{e}^{\dagger}}\Big) +6 {\mu}_{1} \mbox{Tr}\Big({Y_u  Y_{u}^{\dagger}}\Big),\\ 
\beta_{{\mu}_{3}} & = &
-18 g_{2}^{2} {\mu}_{3}  -2 {\mu}_{1} {\sigma}_2  + 4 {\mu}_{3} {\rho}_{1}  -4 {\mu}_{3} {\rho}_{5}  -4 \mu^* {\sigma}_4^*  -4 \mu {\sigma}_4  + 8 {\mu}_{3} {\rho}_{2}  + 8 {\mu}_{3} {\rho}_{4}  -\frac{18}{5} g_{1}^{2} {\mu}_{3},\nn\\\\
\beta_{m^2_{\xi}} & =&
2 \Big(10 m^2_{\xi} {\rho}_{3}  + 2 m^2_{\chi} {\rho}_{5}  + 4 m^2_{\phi} {\sigma}_3  -6 g_{2}^{2} m^2_{\xi}  + 6 m^2_{\chi} {\rho}_{4}  + {\mu}_{1}^{2} + {\mu}_{3}^{2}\Big),\\
\beta_{m^2_{\phi}} & =&
-\frac{9}{10} g_{1}^{2} m^2_{\phi} -\frac{9}{2} g_{2}^{2} m^2_{\phi} +12 {\lambda}_{1} m^2_{\phi} +3 {\mu}_{1}^{2} +6 m^2_{\chi} {\sigma}_1 +3 m^2_{\chi} {\sigma}_2 +6 m^2_{\xi} {\sigma}_3 +12 |\mu|^2 \nonumber \\
&+&6 m^2_{\phi} \mbox{Tr}\Big({Y_d  Y_{d}^{\dagger}}\Big) +2 m^2_{\phi} \mbox{Tr}\Big({Y_e  Y_{e}^{\dagger}}\Big) +6 m^2_{\phi} \mbox{Tr}\Big({Y_u  Y_{u}^{\dagger}}\Big),\eeqa 
\beqa
\beta_{m^2_{\chi}} & =&
-12 g_{2}^{2} m^2_{\chi}  + 12 m^2_{\chi} {\rho}_{2}  + 16 m^2_{\chi} {\rho}_{1}  + 2 m^2_{\xi} {\rho}_{5}  + 2 m^2_{\phi} {\sigma}_2  + 2 {\mu}_{3}^{2}  + 4 m^2_{\phi} {\sigma}_1  + 4 |\mu|^2  \nonumber\quad\quad
\quad \\
&+& 6 m^2_{\xi} {\rho}_{4}  -\frac{18}{5} g_{1}^{2} m^2_{\chi}.
\label{beta_mass}
\eeqa



\begin{thebibliography}{99}
\vspace{-1mm}

\bibitem{ATLAS:higgs}
G.~Aad \textit{et al.} [ATLAS],
Phys. Lett. B \textbf{710} (2012), 49-66
doi:10.1016/j.physletb.2012.02.044
[arXiv:1202.1408 [hep-ex]].

\bibitem{CMS:higgs}
S.~Chatrchyan \textit{et al.} [CMS],
Phys. Lett. B \textbf{710} (2012), 26-48
doi:10.1016/j.physletb.2012.02.064
[arXiv:1202.1488 [hep-ex]].

\bibitem{GM}
H.~Georgi and M.~Machacek,
Nucl. Phys. B \textbf{262} (1985), 463-477
doi:10.1016/0550-3213(85)90325-6.

\bibitem{GM2}
M.~S.~Chanowitz and M.~Golden,
Phys. Lett. B \textbf{165} (1985), 105-108
doi:10.1016/0370-2693(85)90700-2.


\bibitem{Triplet:neutrino1} W. Konetschny and W. Kummer, Phys. Lett. B 70, 433 (1977).
\bibitem{Triplet:neutrino2} T. P. Cheng and L.-F. Li, Phys. Rev. D 22, 2860 (1980).
\bibitem{Triplet:neutrino3} J. Schechter and J. W. F. Valle, Phys. Rev. D 22, 2227 (1980).
\bibitem{Triplet:neutrino4} M. Magg and C. Wetterich, Phys. Lett. B 94, 61 (1980).
\bibitem{Triplet:neutrino5} G. Lazarides, Q. Shafi, and C. Wetterich, Nucl. Phys. B 181, 287 (1981).
\bibitem{Triplet:neutrino6} R. N. Mohapatra and G. Senjanovic, Phys. Rev. D 23, 165 (1981).
\bibitem{GM:neutrino} J. F. Gunion, R. Vega and J. Wudka, Phys. Rev. D 42, 1673 (1990).


\bibitem{GM3}
S.~L.~Chen, A.~Dutta Banik and Z.~K.~Liu,
Nucl. Phys. B \textbf{966} (2021), 115394\\
doi:10.1016/j.nuclphysb.2021.115394
[arXiv:2011.13551 [hep-ph]].



\bibitem{GM52}
C.~W.~Chiang and T.~Yamada,
Phys. Lett. B \textbf{735} (2014), 295-300\\
doi:10.1016/j.physletb.2014.06.048
[arXiv:1404.5182 [hep-ph]].

\bibitem{GM53}
T.~K.~Chen, C.~W.~Chiang, C.~T.~Huang and B.~Q.~Lu,
Phys. Rev. D \textbf{106} (2022) no.5, 055019\\
doi:10.1103/PhysRevD.106.055019
[arXiv:2205.02064 [hep-ph]].

\bibitem{GM54}
R.~Zhou, W.~Cheng, X.~Deng, L.~Bian and Y.~Wu,
JHEP \textbf{01} (2019), 216\\
doi:10.1007/JHEP01(2019)216
[arXiv:1812.06217 [hep-ph]].

\bibitem{GM55}
L.~Bian, H.~K.~Guo, Y.~Wu and R.~Zhou,
Phys. Rev. D \textbf{101} (2020) no.3, 035011\\
doi:10.1103/PhysRevD.101.035011
[arXiv:1906.11664 [hep-ph]].

\bibitem{GM48}
X.~K.~Du, Z.~Li, F.~Wang and Y.~K.~Zhang,
Eur. Phys. J. C \textbf{83} (2023) no.2, 139\\
doi:10.1140/epjc/s10052-023-11297-1
[arXiv:2204.05760 [hep-ph]].

\bibitem{GM49}
P.~Mondal,
Phys. Lett. B \textbf{833} (2022), 137357\\
doi:10.1016/j.physletb.2022.137357
[arXiv:2204.07844 [hep-ph]].

\bibitem{GM50}
T.~K.~Chen, C.~W.~Chiang and K.~Yagyu,
Phys. Rev. D \textbf{106} (2022) no.5, 055035\\
doi:10.1103/PhysRevD.106.055035
[arXiv:2204.12898 [hep-ph]].

\bibitem{GM51}
R.~Ghosh, B.~Mukhopadhyaya and U.~Sarkar,
J. Phys. G \textbf{50} (2023) no.7, 075003\\
doi:10.1088/1361-6471/acd0c8
[arXiv:2205.05041 [hep-ph]].

\bibitem{GM24}
C.~W.~Chiang and K.~Yagyu,
JHEP \textbf{01} (2013), 026\\
doi:10.1007/JHEP01(2013)026
[arXiv:1211.2658 [hep-ph]].

\bibitem{GM25}
S.~Kanemura, M.~Kikuchi and K.~Yagyu,
Phys. Rev. D \textbf{88} (2013), 015020\\
doi:10.1103/PhysRevD.88.015020
[arXiv:1301.7303 [hep-ph]].

\bibitem{GM26}
C.~Englert, E.~Re and M.~Spannowsky,
Phys. Rev. D \textbf{87} (2013) no.9, 095014\\
doi:10.1103/PhysRevD.87.095014
[arXiv:1302.6505 [hep-ph]].

\bibitem{GM27}
C.~Englert, E.~Re and M.~Spannowsky,
Phys. Rev. D \textbf{88} (2013), 035024\\
doi:10.1103/PhysRevD.88.035024
[arXiv:1306.6228 [hep-ph]].

\bibitem{GM28}
S.~I.~Godunov, M.~I.~Vysotsky and E.~V.~Zhemchugov,
J. Exp. Theor. Phys. \textbf{120} (2015) no.3, 369-375\\
doi:10.1134/S1063776115030073
[arXiv:1408.0184 [hep-ph]].

\bibitem{GM29}
C.~W.~Chiang and K.~Tsumura,
JHEP \textbf{04} (2015), 113\\
doi:10.1007/JHEP04(2015)113
[arXiv:1501.04257 [hep-ph]].

\bibitem{GM30}
C.~W.~Chiang, A.~L.~Kuo and T.~Yamada,
JHEP \textbf{01}, 120 (2016)\\
doi:10.1007/JHEP01(2016)120
[arXiv:1511.00865 [hep-ph]].

\bibitem{GM31}
J.~Chang, C.~R.~Chen and C.~W.~Chiang,
JHEP \textbf{03} (2017), 137\\
doi:10.1007/JHEP03(2017)137
[arXiv:1701.06291 [hep-ph]].


\bibitem{GM33}
N.~Ghosh, S.~Ghosh and I.~Saha,
Phys. Rev. D \textbf{101} (2020) no.1, 015029\\
doi:10.1103/PhysRevD.101.015029
[arXiv:1908.00396 [hep-ph]].

\bibitem{GM34}
A.~Ismail, H.~E.~Logan and Y.~Wu,
[arXiv:2003.02272 [hep-ph]].

\bibitem{GM35}
C.~Wang, J.~Q.~Tao, M.~A.~Shahzad, G.~M.~Chen and S.~Gascon-Shotkin,
Chin. Phys. C \textbf{46} (2022) no.8, 083107\\
doi:10.1088/1674-1137/ac6cd3
[arXiv:2204.09198 [hep-ph]].

\bibitem{GM36}
S.~Ghosh,
Int. J. Mod. Phys. A \textbf{39} (2024) no.32, 2450139\\
doi:10.1142/S0217751X24501392
[arXiv:2205.03896 [hep-ph]].


\bibitem{GM37}
M.~Chakraborti, D.~Das, N.~Ghosh, S.~Mukherjee and I.~Saha,
Phys. Rev. D \textbf{109} (2024) no.1, 015016\\
doi:10.1103/PhysRevD.109.015016
[arXiv:2308.02384 [hep-ph]].

\bibitem{GM38}
S.~Ghosh,
LHEP \textbf{2024} (2024), 518\\
doi:10.31526/lhep.2024.518
[arXiv:2311.15405 [hep-ph]].

\bibitem{GM39}
Y.~Zhang, H.~Sun, X.~Luo and W.~Zhang,
Phys. Rev. D \textbf{95} (2017) no.11, 115022\\
doi:10.1103/PhysRevD.95.115022
[arXiv:1706.01490 [hep-ph]].

\bibitem{GM40}
G.~Azuelos, H.~Sun and K.~Wang,
Phys. Rev. D \textbf{97} (2018) no.11, 116005\\
doi:10.1103/PhysRevD.97.116005
[arXiv:1712.07505 [hep-ph]].

\bibitem{GM41}
J.~W.~Zhu, R.~Y.~Zhang, W.~G.~Ma, Q.~Yang, M.~M.~Long and Y.~Jiang,
J. Phys. G \textbf{47} (2020) no.12, 125005\\
doi:10.1088/1361-6471/abaddf.

\bibitem{GM63}
S.~Ghosh,
doi:10.31526/LHEP.2024.518
[arXiv:2311.15405 [hep-ph]].

\bibitem{GM16}
A.~Falkowski, S.~Rychkov and A.~Urbano,
JHEP \textbf{04} (2012), 073\\
doi:10.1007/JHEP04(2012)073
[arXiv:1202.1532 [hep-ph]].

\bibitem{GM17}
S.~Chang, C.~A.~Newby, N.~Raj and C.~Wanotayaroj,
Phys. Rev. D \textbf{86} (2012), 095015\\
doi:10.1103/PhysRevD.86.095015
[arXiv:1207.0493 [hep-ph]].

\bibitem{GM18}
C.~W.~Chiang, A.~L.~Kuo and K.~Yagyu,
JHEP \textbf{10} (2013), 072\\
doi:10.1007/JHEP10(2013)072
[arXiv:1307.7526 [hep-ph]].

\bibitem{GM19}
Y.~Yu, Y.~P.~Bi and J.~F.~Shen,
Phys. Lett. B \textbf{759} (2016), 513-519\\
doi:10.1016/j.physletb.2016.06.014.

\bibitem{GM20}
J.~Cao and Y.~B.~Liu,
Mod. Phys. Lett. A \textbf{31} (2016) no.32, 1650182\\
doi:10.1142/S0217732316501820.

\bibitem{GM21}
H.~Sun, X.~Luo, W.~Wei and T.~Liu,
Phys. Rev. D \textbf{96} (2017) no.9, 095003\\
doi:10.1103/PhysRevD.96.095003
[arXiv:1710.06284 [hep-ph]].

\bibitem{GM22}
J.~Cao, Y.~Q.~Li and Y.~B.~Liu,
Int. J. Mod. Phys. A \textbf{33} (2018) no.11, 1841003\\
doi:10.1142/S0217751X18410038.

\bibitem{GM23}
Q.~Yang, R.~Y.~Zhang, W.~G.~Ma, Y.~Jiang, X.~Z.~Li and H.~Sun,
Phys. Rev. D \textbf{98} (2018) no.5, 055034\\
doi:10.1103/PhysRevD.98.055034
[arXiv:1810.08965 [hep-ph]].

\bibitem{GM42}
A.~Ahriche,
Phys. Rev. D \textbf{110} (2024) no.3, 3\\
doi:10.1103/PhysRevD.110.035010
[arXiv:2312.10484 [hep-ph]].

\bibitem{GM43}
T.~K.~Chen, C.~W.~Chiang, S.~Heinemeyer and G.~Weglein,
Phys. Rev. D \textbf{109} (2024) no.7, 075043\\
doi:10.1103/PhysRevD.109.075043
[arXiv:2312.13239 [hep-ph]].

\bibitem{GM57}
M.~Aoki and S.~Kanemura,
Phys. Rev. D \textbf{77} (2008) no.9, 095009
[erratum: Phys. Rev. D \textbf{89} (2014) no.5, 059902]\\
doi:10.1103/PhysRevD.77.095009
[arXiv:0712.4053 [hep-ph]].

\bibitem{GM58}
K.~Hartling, K.~Kumar and H.~E.~Logan,
Phys. Rev. D \textbf{90} (2014) no.1, 015007\\
doi:10.1103/PhysRevD.90.015007
[arXiv:1404.2640 [hep-ph]].

\bibitem{GM59}
C.~W.~Chiang, G.~Cottin and O.~Eberhardt,
Phys. Rev. D \textbf{99} (2019) no.1, 015001
doi:10.1103/PhysRevD.99.015001
[arXiv:1807.10660 [hep-ph]].

\bibitem{GM60}
D.~Azevedo, P.~Ferreira, H.~E.~Logan and R.~Santos,
JHEP \textbf{03}, 221 (2021)
doi:10.1007/JHEP03(2021)221
[arXiv:positive definite [hep-ph]].

\bibitem{GM61}
Z.~Bairi and A.~Ahriche,
Phys. Rev. D \textbf{108} (2023) no.5, 5
doi:10.1103/PhysRevD.108.055028
[arXiv:2207.00142 [hep-ph]].

\bibitem{GM62}
J.~F.~Gunion, R.~Vega and J.~Wudka,
Phys. Rev. D \textbf{43} (1991), 2322-2336
doi:10.1103/PhysRevD.43.2322.

\bibitem{GM64}
D.~Chowdhury, P.~Mondal and S.~Samanta,
[arXiv:2404.18996 [hep-ph]].

\bibitem{GM56}
K.~Hartling, K.~Kumar and H.~E.~Logan,
Phys. Rev. D \textbf{91} (2015) no.1, 015013
doi:10.1103/PhysRevD.91.015013
[arXiv:1410.5538 [hep-ph]].


\bibitem{VS_SM1}
J.~Elias-Miro, J.~R.~Espinosa, G.~F.~Giudice, G.~Isidori, A.~Riotto and A.~Strumia,
Phys. Lett. B \textbf{709} (2012), 222-228
doi:10.1016/j.physletb.2012.02.013
[arXiv:1112.3022 [hep-ph]].

\bibitem{VS_SM2}
G.~Degrassi, S.~Di Vita, J.~Elias-Miro, J.~R.~Espinosa, G.~F.~Giudice, G.~Isidori and A.~Strumia,
JHEP \textbf{08} (2012), 098
doi:10.1007/JHEP08(2012)098
[arXiv:1205.6497 [hep-ph]].


\bibitem{GM:onelooprho} J. F. Gunion, R. Vega and J. Wudka, Phys. Rev. D 43, 2322 (1991).

\bibitem{Chiang:2017vvo}
C.~W.~Chiang, A.~L.~Kuo and K.~Yagyu,
Phys. Lett. B \textbf{774} (2017), 119-122
doi:10.1016/j.physletb.2017.09.061
[arXiv:1707.04176 [hep-ph]].

\bibitem{Kikuchi:2013gba}
M.~Kikuchi,
Nuovo Cim. C \textbf{037} (2014) no.02, 125-130
doi:10.1393/ncc/i2014-11745-y
[arXiv:1312.7641 [hep-ph]].

\bibitem{Chiang:2018xpl}
C.~W.~Chiang, A.~L.~Kuo and K.~Yagyu,
Phys. Rev. D \textbf{98} (2018) no.1, 013008
doi:10.1103/PhysRevD.98.013008
[arXiv:1804.02633 [hep-ph]].

\bibitem{GM44}
S.~Blasi, S.~De Curtis and K.~Yagyu,
Phys. Rev. D \textbf{96}, no.1, 015001 (2017)
doi:10.1103/PhysRevD.96.015001
[arXiv:1704.08512 [hep-ph]].

\bibitem{GM45}
B.~Keeshan, H.~E.~Logan and T.~Pilkington,
Phys. Rev. D \textbf{102} (2020) no.1, 015001
doi:10.1103/PhysRevD.102.015001
[arXiv:1807.11511 [hep-ph]].

\bibitem{GM32}
B.~Li, Z.~L.~Han and Y.~Liao,
JHEP \textbf{02} (2018), 007
doi:10.1007/JHEP02(2018)007
[arXiv:1710.00184 [hep-ph]].

\bibitem{Aoki:2007ah}
M.~Aoki and S.~Kanemura,
Phys.\ Rev.\ D {\bf 77}, 095009 (2008)
[arXiv:0712.4053 [hep-ph]];
erratum Phys.\ Rev.\ D {\bf 89}, 059902 (2014).

\bibitem{Moultaka:2020dmb}
G.~Moultaka and M.~C.~Peyran\`ere,
Phys. Rev. D \textbf{103} (2021) no.11, 115006
doi:10.1103/PhysRevD.103.115006
[arXiv:2012.13947 [hep-ph]].

\bibitem{cw}
S.~R.~Coleman and E.~J.~Weinberg,
Phys. Rev. D \textbf{7} (1973), 1888-1910
doi:10.1103/PhysRevD.7.1888.

\bibitem{sher}
M.~Sher,
Phys. Rept. \textbf{179} (1989), 273-418
doi:10.1016/0370-1573(89)90061-6.

\bibitem{Kastening}
B.~M.~Kastening,
Phys. Lett. B \textbf{283} (1992), 287-292
doi:10.1016/0370-2693(92)90021-U.

\bibitem{Machacek:1983tz}
M.~E.~Machacek and M.~T.~Vaughn,
Nucl. Phys. B \textbf{222} (1983), 83-103
doi:10.1016/0550-3213(83)90610-7.

\bibitem{Machacek:1983fi}
M.~E.~Machacek and M.~T.~Vaughn,
Nucl. Phys. B \textbf{236} (1984), 221-232
doi:10.1016/0550-3213(84)90533-9.

\bibitem{Machacek:1984zw}
M.~E.~Machacek and M.~T.~Vaughn,
Nucl. Phys. B \textbf{249} (1985), 70-92
doi:10.1016/0550-3213(85)90040-9.

\bibitem{SARAH1}
F.~Staub,
Comput. Phys. Commun. \textbf{185} (2014), 1773-1790
doi:10.1016/j.cpc.2014.02.018
[arXiv:1309.7223 [hep-ph]].

\bibitem{SARAH2}
F.~Staub,
Comput. Phys. Commun. \textbf{184} (2013), 1792-1809
doi:10.1016/j.cpc.2013.02.019
[arXiv:1207.0906 [hep-ph]].

\bibitem{SARAH3}
F.~Staub,
Comput. Phys. Commun. \textbf{182} (2011), 808-833
doi:10.1016/j.cpc.2010.11.030
[arXiv:1002.0840 [hep-ph]].

\bibitem{SARAH4}
F.~Staub,
Comput. Phys. Commun. \textbf{181} (2010), 1077-1086
doi:10.1016/j.cpc.2010.01.011
[arXiv:0909.2863 [hep-ph]].

\bibitem{SARAH5}
F.~Staub,
[arXiv:0806.0538 [hep-ph]].

\bibitem{SPheno1}
W.~Porod,
Comput. Phys. Commun. \textbf{153} (2003), 275-315
doi:10.1016/S0010-4655(03)00222-4
[arXiv:hep-ph/0301101 [hep-ph]].

\bibitem{SPheno2}
W.~Porod and F.~Staub,
Comput. Phys. Commun. \textbf{183} (2012), 2458-2469
doi:10.1016/j.cpc.2012.05.021
[arXiv:1104.1573 [hep-ph]].

\bibitem{GKZ}
I. Gelfand, M. Kapranov and A. Zelevinsky, Discriminants, resultants and multidimensional determinants, Modern Birkhauser Classics (1994).

\bibitem{Fernando}
Fernando Cukierman, A CRITERION FOR POSITIVE POLYNOMIALS, arXiv:math/0312469 [math.AG].

\bibitem{Miao}
Miao Yuan, Li Chunwen. Definiteness of Multi-Variable Homogeneous Polynomial. ACTA AUTOMATICA SINICA, 1998, 24(4): 539-542.

\bibitem{Miao1}
Li Chunwen, Miao Yuan, Miao Qinghai.  A method to judge the stability of dynamical system. In Proceeding of YAC'95, IFAC, 1995, Pergamon Press, 101-106.

\bibitem{Bose}
N. K. Bose, Inner Algorithm to Test for Positive Definiteness of Arbitrary Binary Form, IEEE T-AC, vol. 19, 169-170 (1975).
    
\bibitem{Bose1}
N. K. Bose, Test for Lynapunov Stability by Rational Operations, IEEE T-AC, vol. 19, 700-702 (1975).

\bibitem{Antusch:2013jca}
S.~Antusch and V.~Maurer,
JHEP \textbf{11} (2013), 115
doi:10.1007/JHEP11(2013)115
[arXiv:1306.6879 [hep-ph]].

\bibitem{ATLAS:kappaWZ}
G. Aad et al. [ATLAS Collaboration], Nature 607, 52-59 (2022).

\bibitem{CMS:kappaWZ}
S. Chatrachyan et al. [CMS Collaboration], Nature 607, 60-68(2022).

\bibitem{ParticleDataGroup:2022pth}
R.~L.~Workman \textit{et al.} [Particle Data Group],
Progress of Theoretical and Experimental Physics, PTEP\textbf{2022} (2022), 083C01
doi:10.1093/ptep/ptac097.

\bibitem{B-physics}
V.~Khachatryan \textit{et al.} [CMS and LHCb],
Nature \textbf{522} (2015), 68-72
doi:10.1038/nature14474
[arXiv:1411.4413 [hep-ex]].

\bibitem{HiggsBounds1}
P.~Bechtle, O.~Brein, S.~Heinemeyer, G.~Weiglein and K.~E.~Williams,
Comput. Phys. Commun. \textbf{181} (2010), 138-167
doi:10.1016/j.cpc.2009.09.003
[arXiv:0811.4169 [hep-ph]].

\bibitem{HiggsBounds2}
P.~Bechtle, O.~Brein, S.~Heinemeyer, G.~Weiglein and K.~E.~Williams,
Comput. Phys. Commun. \textbf{182} (2011), 2605-2631
doi:10.1016/j.cpc.2011.07.015
[arXiv:1102.1898 [hep-ph]].

\bibitem{HiggsBounds3}
P. Bechtle, O. Brein, S. Heinemeyer, O. Stal, T. Stefaniak, G. Weiglein and K. E. Williams,
Eur. Phys. J. C 74, 2693 (2014), arXiv:1311.0055.

\bibitem{HiggsBounds4}
P.~Bechtle, D.~Dercks, S.~Heinemeyer, T.~Klingl, T.~Stefaniak, G.~Weiglein and J.~Wittbrodt,
Eur. Phys. J. C \textbf{80} (2020) no.12, 1211
doi:10.1140/epjc/s10052-020-08557-9
[arXiv:2006.06007 [hep-ph]].

\bibitem{HiggsSignals1}
P. Bechtle, S. Heinemeyer, O. Stal, T. Stefaniak and G. Weiglein, Eur. Phys. J. C 74, 2711 (2014),
  arXiv:1305.1933.

\bibitem{HiggsSignals2}
P. Bechtle, S. Heinemeyer, O. Stal, T. Stefaniak and G. Weiglein, JHEP 1411, 039 (2014), [arXiv:1403.1582].

\bibitem{HiggsSignals3}
P.~Bechtle, S.~Heinemeyer, T.~Klingl, T.~Stefaniak, G.~Weiglein and J.~Wittbrodt,
Eur. Phys. J. C \textbf{81} (2021) no.2, 145
doi:10.1140/epjc/s10052-021-08942-y
[arXiv:2012.09197 [hep-ph]].

\bibitem{HiggsTools}
H.~Bahl, T.~Biek\"otter, S.~Heinemeyer, C.~Li, S.~Paasch, G.~Weiglein and J.~Wittbrodt,
Comput. Phys. Commun. \textbf{291} (2023), 108803
doi:10.1016/j.cpc.2023.108803
[arXiv:2210.09332 [hep-ph]].

\bibitem{GMCal}
K.~Hartling, K.~Kumar and H.~E.~Logan,
[arXiv:1412.7387 [hep-ph]].

\bibitem{HL-LHC1}
I.~Zurbano Fernandez, M.~Zobov, A.~Zlobin, F.~Zimmermann, M.~Zerlauth, C.~Zanoni, C.~Zannini, O.~Zagorodnova, I.~Zacharov and M.~Yu, \textit{et al.}
CERN, 2020,
ISBN 978-92-9083-586-8, 978-92-9083-587-5
doi:10.23731/CYRM-2020-0010

\bibitem{HL-LHC2}
R.~Aaij \textit{et al.} [LHCb],
[arXiv:1808.08865 [hep-ex]].

\bibitem{CEPC1}
H.~Cheng \textit{et al.} [CEPC Physics Study Group],
[arXiv:2205.08553 [hep-ph]].

\bibitem{CEPC2}
W.~Abdallah \textit{et al.} [CEPC Study Group],
Radiat. Detect. Technol. Methods \textbf{8} (2024) no.1, 1-1105
doi:10.1007/s41605-024-00463-y
[arXiv:2312.14363 [physics.acc-ph]].

\bibitem{CEPC3}
X.~Ai, W.~Altmannshofer, P.~Athron, X.~Bai, L.~Calibbi, L.~Cao, Y.~Che, C.~Chen, J.~Y.~Chen and L.~Chen, \textit{et al.}
[arXiv:2412.19743 [hep-ex]].


\bibitem{oneloop:RGE}
T. P. Cheng, E. Eichten, and L. F. Li, "Higgs Phenomena in
 asymptotically free gauge theories," Phys. Rev. D 9, 2259 (1974).

\bibitem{Kundu:2021pcg}
A.~Kundu, P.~Mondal and P.~B.~Pal,
Phys. Rev. D \textbf{105} (2022) no.11, 115026
doi:10.1103/PhysRevD.105.115026
[arXiv:2111.14195 [hep-ph]].

\end{thebibliography}
\end{document}